\documentclass[12pt,a4paper]{article}
\usepackage[T1]{fontenc}
\usepackage[latin1]{inputenc}
\usepackage{color}
\usepackage[dvipsnames]{xcolor}	
\usepackage{enumitem}
\usepackage{graphicx}
\usepackage{booktabs}
\usepackage{multicol}
\usepackage{multirow}
\usepackage{float} 

\usepackage{amsfonts}
\usepackage{amsmath}
\usepackage{amssymb}
\usepackage{theorem}
\usepackage{mathtools}
\usepackage{empheq}
\usepackage{upgreek}

\theoremstyle{plain}
\theorembodyfont{\slshape}
\newtheorem{satz}{Theorem}[section]
\newtheorem{lemma}[satz]{Lemma}

\newtheorem{kor}[satz]{Corollary}

\theorembodyfont{\upshape\small}
\newtheorem{bsp}[satz]{Example}
\newtheorem{bem}[satz]{Remark}

\theorembodyfont{\ttfamily}

\numberwithin{equation}{section}
\newcommand{\ba}{\begin{equation}}
\newcommand{\ea}{\end{equation}}

\newcommand{\Z}{\mbox{\boldmath $Z$}}
\newcommand{\D}{\mbox{\boldmath $D$}}
\newcommand{\Hes}{\mbox{\boldmath $H$}}
\newcommand{\bmu}{\mbox{\boldmath $\mu$}}

\newcommand{\bSigma}{\mbox{\boldmath $\Sigma$}}

\newcommand*\diff{\mathop{}\!\mathrm{d}}

\newcommand{\Exp}{\textup{Exp}}
\newcommand{\IG}{\textup{IG}}

\newcommand{\gam}{\textup{Gamma}}
\newcommand{\nb}{\textup{NB}}
\newcommand{\norm}{\textup{N}}

\newcommand{\bbn}{\mathbb{N}}
\newcommand{\bbr}{\mathbb{R}}

\newcommand{\e}{\mathbb{E}}
\newcommand{\V}{\mathbb{V}}
\newcommand{\Pro}{\mathbb{P}}
\newcommand{\pgf}{\text{pgf}\,}

\newcommand{\iid}{i.\,i.\,d.}
\newcommand{\ie}{i.\,e., }
\newcommand{\eg}{e.\,g., }

\usepackage{dsfont}

\usepackage{natbib}

\usepackage{hyperref}
\hypersetup{
	colorlinks,
	linkcolor={red!90!black}, 
	citecolor={green!50!black}, 
	urlcolor={black}
}

\begin{document}



\parindent 0cm

\title{Generalized Moment Estimators based on Stein Identities}
\author{
Simon Nik\thanks{
Helmut Schmidt University, Department of Mathematics and Statistics, Hamburg, Germany.}
\and 
Christian H.\ Wei\ss\footnotemark[1]\ \thanks{Corresponding author. E-Mail: \href{mailto:weissc@hsu-hh.de}{\nolinkurl{weissc@hsu-hh.de}}. ORCID: \href{https://orcid.org/0000-0001-8739-6631}{\nolinkurl{0000-0001-8739-6631}}.}
}

\maketitle

\begin{abstract}
\noindent
For parameter estimation of continuous and discrete distributions, we propose a generalization of the method of moments (MM), where Stein identities are utilized for improved estimation performance. The construction of these Stein-type MM-estimators makes use of a weight function as implied by an appropriate form of the Stein identity. Our general approach as well as potential benefits thereof are first illustrated by the simple example of the exponential distribution. Afterward, we investigate the more sophisticated two-parameter inverse Gaussian distribution and the two-parameter negative-binomial distribution in great detail, together with illustrative real-world data examples. Given an appropriate choice of the respective weight functions, their Stein-MM estimators, which are defined by simple closed-form formulas and allow for closed-form asymptotic computations, exhibit a better performance regarding bias and mean squared error than competing estimators.

\end{abstract}

%
\section{Introduction}\label{Introduction}
%
For many parametric distributions, so-called Stein identities are available, which rely on moments of functional expressions of a corresponding random variable. These identities are named after Charles Stein, who developed the idea of uniquely characterizing a certain distribution family by such a moment identity \citep[see][]{stein72,stein86}. Many examples of both continuous and discrete distributions together with their Stein characterizations can be found in \citet{Stein04,sudheesh09,sudheesh12,landsman16,weissaleksandrov22,anastasiou23} and the references therein. Stein identities are not a mere tool of probability theory. During the last years, there was also a lot of research activity on statistical applications of Stein identities, for example to goodness-of-fit (GoF) tests \citep{betsch22,weissetal23} and control charts \citep{weiss23}, among others. 
In the present article, however, another application of Stein identities is investigated and exemplified, namely to the parameter estimation of continuous or discrete distributions. The idea to construct generalized types of method-of-moments (MM) estimators based on an appropriate type of Stein identity plus weighting function, referred to as Stein-MM estimators, was first explored in some applications by \citet{arnold01} and \citet{wang23}. Recently, \citet{ebner23} discussed the Stein-MM approach in a much broader way, and also the present article provides a comprehensive treatment of Stein-MM estimators for various distributions. 
The main motivation for considering Stein-MM estimation is that the weighting function might be chosen in such a way that the resulting estimator shows better properties (\eg a reduced bias or mean squared error (MSE)) than the default MM~estimator or other existing estimators. Despite the additional flexibility offered by the weighting function, the Stein-MM estimators are computed from simple closed-form expressions, and consistency and asymptotic normality are easily established, also see \citet{ebner23}. 

\smallskip
In what follows, we apply the proposed Stein-MM estimation to three different distribution families. We start with the illustrative example of the exponential (Exp) distribution in Section~\ref{Stein Estimation of Exponential Distribution}. This simple one-parameter distribution mainly serves to demonstrate the general approach for deriving the Stein-MM estimator and its asymptotics, and it also indicates the potential benefits of using the Stein-MM approach for parameter estimation. 
Afterward in Section~\ref{Stein Estimation of Inverse-Gaussian Distribution}, we examine a more sophisticated type of continuous distribution, namely the two-parameter inverse Gaussian (IG) distribution. In Section~\ref{Stein Estimation of Negative-binomial Distribution}, we then turn to a discrete distribution family, namely the two-parameter negative-binomial (NB) distribution. Illustrative real-world data examples are also presented in Sections~\ref{Stein Estimation of Inverse-Gaussian Distribution}--\ref{Stein Estimation of Negative-binomial Distribution}. 
Note that neither the exponential distribution nor any discrete distribution have been considered by \citet{ebner23}, and their approach to the Stein-MM estimation of the IG-distribution differs from the one proposed here, see the details below. Also \citet{arnold01,wang23} did not discuss any of the aforementioned distributions. 
Finally, we conclude in Section~\ref{Conclusions} and outline topics for future research.

%
\section{Stein Estimation of Exponential Distribution}
\label{Stein Estimation of Exponential Distribution}
%
The exponential distribution is the most well-known lifetime distribution, which is characterized by the property of being memory-less. It has positive support and depends on the parameter $\lambda>0$, where its probability density function (pdf) is given by $\phi(x)=\lambda e^{-\lambda x}$ for $x > 0 $ and zero otherwise. 
A detailed survey about the properties of and estimators for the $\Exp(\lambda)$-distribution can be found in \citet[Chapter 19]{Johnson95}. 
Given the independent and identically distributed (\iid) sample $X_1,\ldots,X_n$ with $ X_i\sim \Exp(\lambda) $ for $ i=1,\ldots,n $, the default estimator of~$\lambda>0$, which is an MM estimator and the maximum likelihood (ML) estimator at the same time, is given by $\hat{\lambda} = 1/\overline{X}$, where $ \overline{X}=\tfrac{1}{n}\sum_{i=1}^{n} X_i$ denotes the sample mean. This estimator is known to neither be unbiased, nor to be optimal in terms of the MSE, see \citet{elfessi01}.
To derive a generalized MM~estimator with perhaps improved bias or MSE properties, we consider the exponential Stein identity according to \citet[Example 1.6]{Stein04}, which states that
\begin{align}
\label{Lionel}
	X\sim\Exp(\lambda) \quad \text{iff} \quad \e[f'(X)]=\lambda\,\e[f(X)]
\end{align}
for any piecewise differentiable function $ f $ with $ f(0)=0 $ such that $ \e\big[\vert f'(X)\vert\big] $, $ \e\big[\vert f(X)\vert\big] $ exist. 
Solving \eqref{Lionel} in~$\lambda$ and using the sample moments $ \overline{f'(X)}=\tfrac{1}{n}\sum_{i=1}^{n} f'(X_i)$ and $ \overline{f(X)}=\tfrac{1}{n}\sum_{i=1}^{n} f(X_i)$ instead of the population moments, the class of Stein-MM estimators for $ \lambda $ is obtained as
\begin{align}
\label{ExpSteinMM}
	\hat{\lambda}_{f,\Exp}	= \frac{\, \overline{f'(X)} \,}{\overline{f(X)}}.
\end{align}
Note that the choice $f(x)=x$ leads to the default estimator $\hat{\lambda} = 1/\overline{X}$. Generally, $f(x) \not= x$ might be interpreted as a kind of weighting function, which assigns different weights to large or low values of~$X$ than the identity function does. 
For deriving the asymptotic distribution of the general Stein-MM estimator~$\hat{\lambda}_{f,\Exp}$, we first define the vectors $\Z_i $ with $ i=1,\ldots,n $ as 
\begin{align}
\label{ExpZi}
	\Z_i\coloneqq\big(f'(X_i),f(X_i)\big)^\top.
\end{align}
Their mean equals
\begin{align}
\label{ExpZiMean}
	\bmu_Z\coloneqq\e[\Z_i]=
	\begin{pmatrix}
		\e[f'(X_i)] \\
		\e[f(X_i)]
	\end{pmatrix} \overset{\eqref{Lionel}}{=}\mu_f(0,1,0)\begin{pmatrix}
		\lambda \\
		1
	\end{pmatrix},
\end{align}
where we define $ \mu_f(k,l,m)\coloneqq \e[X^k\cdot f(X)^l\cdot f'(X)^m]$ for any $k,l,m\in\bbn_0=\{0,1,\ldots\}$. Then, the following central limit theorem (CLT) holds.

\begin{satz}
\label{CLT_Exp}
If $X_1,\ldots,X_n$ are \iid\ according to $ \Exp(\lambda) $, then the sample mean~$\overline{\Z}$ of $\Z_1,\ldots,\Z_n$ according to \eqref{ExpZi} is asymptotically normally distributed as
	\begin{align*}
		\sqrt{n}\big(\overline{\Z}-\bmu_Z\big) \ \xrightarrow{\text{d}}\ \norm\big(\mathbf{0}, \bSigma\big)\quad \text{with}\quad \bSigma=(\sigma_{ij})_{i,j=1,2},
	\end{align*}
	where $ \norm(\mathbf{0}, \bSigma) $ denotes the multivariate normal distribution, and where the covariances are given as 
	\begin{align*}
		\sigma_{11}&=\mu_f(0,0,2)-\lambda^2\cdot\mu_f^2(0,1,0),\\
		\sigma_{22}&=\mu_f(0,2,0)-\mu_f^2(0,1,0),\\
		\sigma_{12}&=\tfrac{\lambda}{2}\big(\sigma_{22}-\mu_f^2(0,1,0)\big) = \lambda\big(\sigma_{22}-\tfrac{1}{2}\mu_f(0,2,0)\big).
	\end{align*}
\end{satz}
The proof of Theorem~\ref{CLT_Exp} is provided by Appendix~\ref{Proof of Theorem CLT_Exp}.
In the second step of deriving the asymptotics of~$\hat{\lambda}_{f,\Exp}$, we define the function $ g(u,v)\coloneqq \frac{u}{v} $. Then, $ \hat{\lambda}_{f,\Exp}=g(\overline{\Z}) $ and $ \lambda=g(\bmu_Z) $. Applying the Delta method \citep{serfling80} to Theorem~\ref{CLT_Exp}, the following result follows.

\begin{satz}
\label{Diego}
If $X_1,\ldots,X_n$ are \iid\ according to $ \Exp(\lambda) $, then  $ \hat{\lambda}_{f,\Exp}$ is asymptotically normally distributed, where the asymptotic variance and bias are given by 
	\begin{align*}
		\sigma_{f,\Exp}^2=\frac{1}{n}\cdot\frac{\mu_f(0,0,2)}{\mu_f^2(0,1,0)}, \qquad
		\mathbb{B}_{f,\Exp}=\frac{\lambda}{2n}\cdot\frac{\mu_f(0,2,0)}{\mu_f^2(0,1,0)}.
	\end{align*}
\end{satz}
The proof of Theorem~\ref{Diego} is provided by Appendix~\ref{Proof of Theorem Diego}. Note that the moments $\mu_f(k,l,m)$ involved in Theorems~\ref{CLT_Exp} and~\ref{Diego} can sometimes be derived explicitly, see the subsequent examples, while they can be computed by using numerical integration otherwise.

\smallskip
After having derived the asymptotic variance and bias without explicitly specifying the function~$f$, let us now consider the special case $f_a: [0;\infty)\to [0;\infty)$, $ f_a(x)=x^a $, as our first illustrative example (where $a=1$ leads to the default estimator $\hat{\lambda} = 1/\overline{X}$). Here, we have to restrict to $a>0$ to ensure that the condition $ f(0)=0 $ in \eqref{Lionel} holds. Using that
\begin{align}
\label{ExpMomXa}
	\e[X^a] = \frac{\Gamma(a+1)}{\lambda^a}
	\quad\text{for } X\sim\Exp(\lambda) \text{ if } a>-1,
\end{align}
the following corollary to Theorem~\ref{Diego} is derived.

\begin{kor}
\label{korExpAsymXa}
	Let $X_1,\ldots,X_n$ be \iid\ according to $ \Exp(\lambda) $, and let $ f_a(x)= x^a $ with $ a>\tfrac{1}{2} $. Then, $ \hat{\lambda}_{f_a,\Exp}$ is asymptotically normally distributed, where the asymptotic variance and bias are given by 
	\begin{align*}
		\sigma_{f_a,\Exp}^2=\frac{\lambda^2}{n}\binom{2(a-1)}{a-1}, \qquad
		\mathbb{B}_{f_a,\Exp}=\frac{\lambda}{2n}\binom{2a}{a}.
	\end{align*}
Furthermore, the MSE equals 
\begin{align*}
	\text{MSE}_{f_a,\Exp}&=	\sigma_{f_a,\Exp}^2+\mathbb{B}_{f_a,\Exp}^2=\lambda^2\bigg[\frac{1}{n}\binom{2(a-1)}{a-1}+\frac{1}{4 n^2}\binom{2a}{a}^2\, \bigg].
\end{align*}
\end{kor}
The proof of \eqref{ExpMomXa} and Corollary~\ref{korExpAsymXa} is provided by Appendix~\ref{Proof of Corollary korExpAsymXa}. Note that in Corollary~\ref{korExpAsymXa}, $\binom{r}{s}$ denotes the generalized binomial coefficient given by $\Gamma(r+1)/\Gamma(s+1)/\Gamma(r-s+1)$.

\begin{figure}[t]
	\center\footnotesize
	(a)\hspace{-3ex}\includegraphics[scale=0.5, viewport=0 45 335 305, clip=
	]{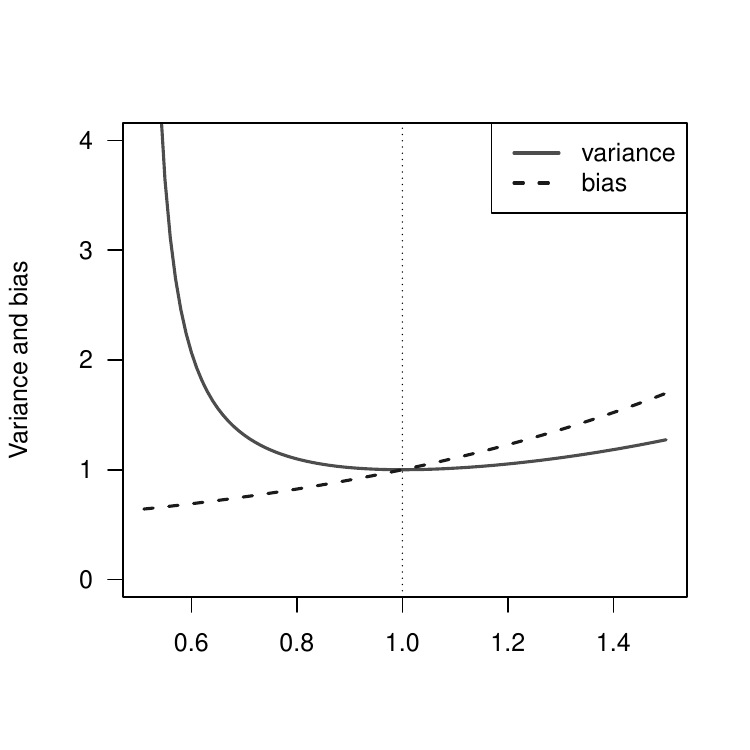}\hspace{-1ex}$a$
	\qquad
	(b)\hspace{-3ex}\includegraphics[scale=0.5, viewport=0 45 335 305, clip=
	]{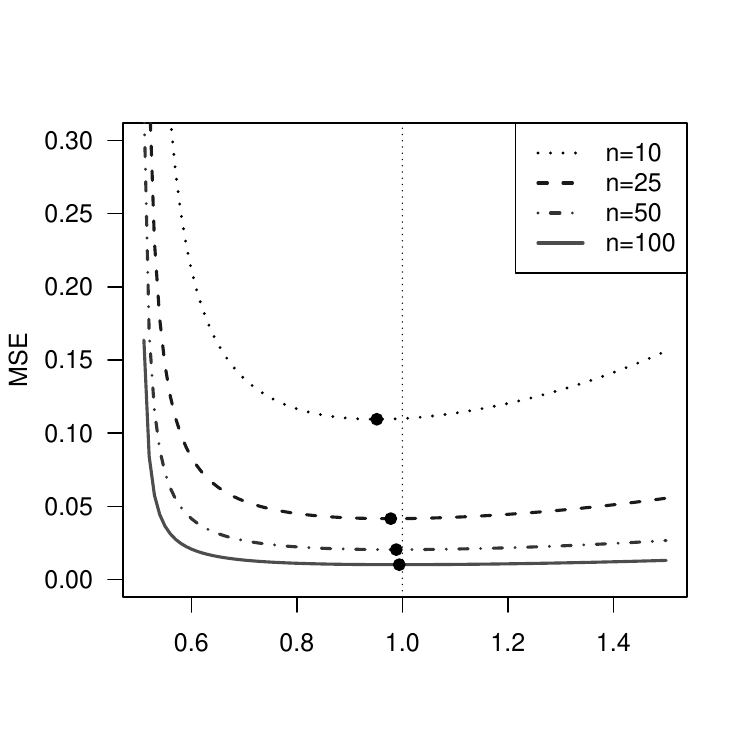}\hspace{-1ex}$a$
	\caption{Plot of (a) $n\,\sigma^2 _{f_a,\Exp} $ and $ n\,\mathbb{B}_{f_a,\Exp} $, and (b) MSE$ _{f_a,\Exp} $ for $ a\in (0.5;1.5) $ and $\lambda=1$. Points indicate minimal MSE values. Dotted line at $a=1$ corresponds to default estimator $\hat{\lambda} = 1/\overline{X}$.}
	\label{VarBias_Exp_xa}
\end{figure}

\smallskip
In Figure \ref{VarBias_Exp_xa}\,(a), the asymptotic variance and bias of $ \hat{\lambda}_{f_a,\Exp}$ according to Corollary~\ref{korExpAsymXa} are presented. While the variance is minimal for $a=1$ (\ie for the ordinary MM and ML estimator), the bias decreases with decreasing~$ a $ (\ie bias reductions are achieved for sublinear choices of $f_a(x)=x^a$). Hence, an MSE-optimal choice of $f_a(x)$ is obtained for some $a\in (0.5;1)$. This is illustrated by Figure \ref{VarBias_Exp_xa}\,(b), where the MSE of Corollary~\ref{korExpAsymXa} is presented for different sample sizes $ n\in \{10,25,50,100\} $. The corresponding optimal values of~$a$ are determined by numerical optimization as 0.952, 0.978, 0.988, and 0.994, respectively. As a result, especially for small $ n $, we have a reduction of the MSE (and of the bias as well) if using a ``true'' Stein-MM estimator (\ie with $a\not=1$). Certainly, if the focus is mainly on bias reduction, then an even smaller choice of~$a$ would be beneficial.

\smallskip
As a second illustrative example, let us consider the functions $ f_u(x)=1-u^x $ with $ u\in(0;1) $, which are again sublinear, but this time also bounded from above by one. Again, we can derive a corollary to Theorem~\ref{Diego}, this time by using the moment formula
\begin{align}
\label{ExpMomuX}
	\e\big[u^X\big]=\frac{\lambda}{\lambda-\ln(u)}
	\quad\text{for } X\sim\Exp(\lambda) \text{ if } u\in(0;1).
\end{align}

\begin{kor}
\label{korExpAsymuX}
Let $X_1,\ldots,X_n$ be \iid\ according to $ \Exp(\lambda) $, and let $ f_u(x)=1- u^x $ with $ u\in(0;1) $. Then, $ \hat{\lambda}_{f_u,\Exp}$ is asymptotically normally distributed, where the asymptotic variance and bias are given by 
	\begin{align*}
		\sigma_{f_u,\Exp}^2=\frac{\lambda}{n}\frac{(\lambda-\ln(u))^2}{\lambda-2\ln(u)}, \qquad
		\mathbb{B}_{f_u,\Exp}=\frac{	\sigma_{f_u,\Exp}^2}{\lambda-\ln(u)}.
	\end{align*}
Furthermore, the MSE equals 
\begin{align*}
	\text{MSE}_{f_u,\Exp}&=\Big(1+\tfrac{\lambda}{n(\lambda-2\ln(u))}\Big)\cdot\sigma_{f_u,\Exp}^2.
\end{align*}
\end{kor}
The proof of \eqref{ExpMomuX} and Corollary~\ref{korExpAsymuX} is provided by Appendix~\ref{Proof of Corollary korExpAsymuX}. 

\begin{figure}[t]
	\center\footnotesize
	(a)\hspace{-3ex}\includegraphics[scale=0.5, viewport=0 45 335 305, clip=
	]{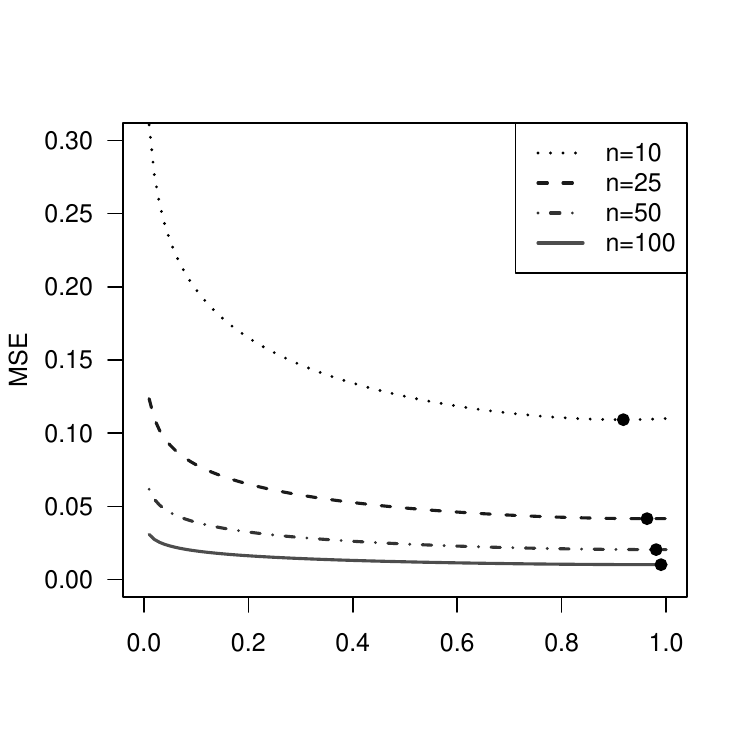}$u$
	\qquad
	(b)\hspace{-3ex}\includegraphics[scale=0.5, viewport=0 45 335 305, clip=
	]{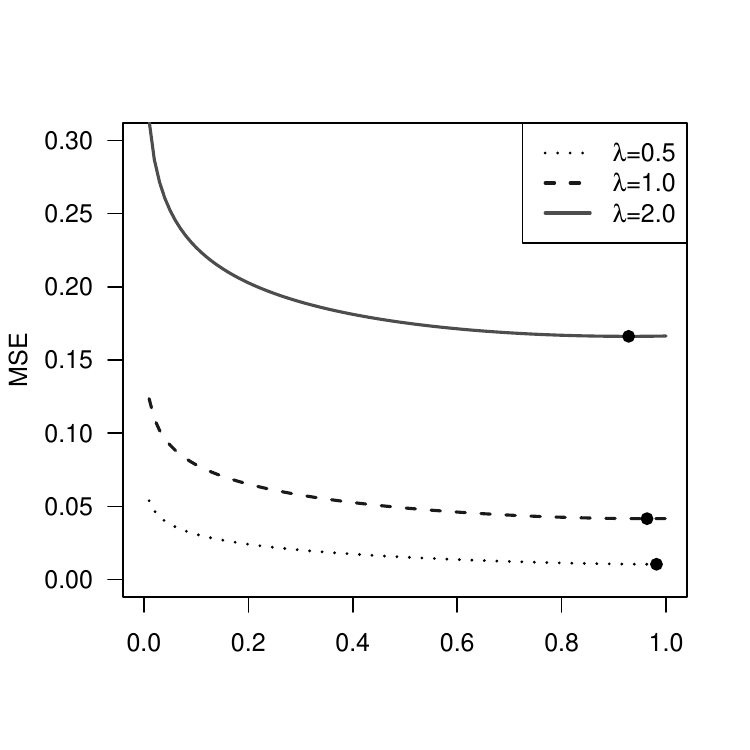}$u$
	\caption{Plot of MSE$ _{f_u,\Exp} $ for $ u\in (0;1) $, for (a) different~$n$ and $\lambda=1$, and (b) different~$\lambda$ and $n=25$. Points indicate minimal MSE values.}
	\label{MSE_Exp_ux}
\end{figure}

\smallskip
This time, the variance decreases for increasing~$u$, whereas the bias decreases for decreasing~$u$. As a consequence, an MSE-optimal choice is expected for some~$u$ inside the interval $(0;1)$. 
This is illustrated by Figure~\ref{MSE_Exp_ux}\,(a), where the minima for $ n\in \{10,25,50,100\} $, given that $\lambda=1$, are attained for $u\approx 0.918$, $0.963$, $0.981$, and $0.990$, respectively. The major difference between the two types of weighting functions in Corollaries~\ref{korExpAsymXa} and~\ref{korExpAsymuX} is given by the role of~$\lambda$ within the expression for the MSE. For~$f_a(x)$ in Corollary~\ref{korExpAsymXa}, $\lambda$ occurs as a simple factor such that the optimal choice for~$a$ is the same across different~$\lambda$. Hence, the optimal~$a$ is simply a function of the sample size~$n$, which is very attractive for applications in practice. For~$f_u(x)$ in Corollary~\ref{korExpAsymuX}, by contrast, the MSE depends in a more sophisticated way on~$\lambda$, and the optimal~$u$ differs for different~$\lambda$ as illustrated by Figure~\ref{MSE_Exp_ux}\,(b). Thus, if one wants to use the weighting function~$f_u(x)$ in practice, a two-step procedure appears reasonable, where an initial estimate is computed via $\hat{\lambda} = 1/\overline{X}$, which is then refined by $ \hat{\lambda}_{f_u,\Exp}$ with~$u$ being determined by plugging-in~$\hat{\lambda}$ instead of~$\lambda$ (also see Section~2.2 in \citet{ebner23} for an analogous idea). 

\medskip
We conclude this section by pointing out two further application scenarios for the use of Stein-MM estimators~$\hat{\lambda}_{f,\Exp}$. First, in analogy to recent Stein-based GoF-tests\label{GoF-test} such as in \citet{weissetal23}, $\hat{\lambda}_{f,\Exp}$ might be used for GoF-applications. More precisely, the idea could be to select a set $\{f_1,\ldots, f_K\}$ of weighting functions, and to compute $\hat{\lambda}_{f_k}$ for all $k=1,\ldots,K$. As any $\hat{\lambda}_{f_k}$ is a consistent estimator of~$\lambda$ according to Theorem~\ref{Diego}, the obtained values $\{\hat{\lambda}_{f_1},\ldots,\hat{\lambda}_{f_K}\}$ should vary closely around~$\lambda$. For other continuous distributions with positive support, such as the IG-distribution considered in the next Section~\ref{Stein Estimation of Inverse-Gaussian Distribution}, we cannot expect that $\overline{f'(X)} / \overline{f(X)}$ has an (asymptotically) unique mean for different~$f$, see Remark~\ref{bemExpGoFtest}, so a larger variation among the values in $\{\hat{\lambda}_{f_1},\ldots,\hat{\lambda}_{f_K}\}$ is expected. Such a discrepancy in variation might give rise for a formal exponential GoF-test. But as the focus of this article is on parameter estimation, we postpone a detailed investigation of this GoF-application to future research.

\begin{table}[t]
\centering\small
\caption{Empirical bias and MSE of $\hat{\lambda}_{f,\Exp}$ from $10^5$ simulated \iid\ samples $X_1,\ldots,X_n\sim \Exp(1)$, where $\lceil 0.1\cdot n\rceil$ observations randomly selected for additive outlier ``$+5$''.}
\label{tabOutliers}

\smallskip
\resizebox{\linewidth}{!}{
\begin{tabular}{l|cccc|cccc}
\toprule
 & \multicolumn{4}{l|}{Bias} & \multicolumn{4}{l}{MSE} \\
$n\ \setminus\ f$ & $x^{0.9}$ & $1-0.9^x$ & $\ln(1+x)$ & $x$ & $x^{0.9}$ & $1-0.9^x$ & $\ln(1+x)$ & $x$ \\
\midrule
10 & -0.280 & -0.277 & -0.206 & -0.304 & 0.107 & 0.105 & 0.100 & 0.114 \\
25 & -0.343 & -0.341 & -0.271 & -0.366 & 0.126 & 0.125 & 0.091 & 0.140 \\
50 & -0.305 & -0.303 & -0.238 & -0.328 & 0.098 & 0.097 & 0.067 & 0.111 \\
100 & -0.308 & -0.306 & -0.241 & -0.330 & 0.098 & 0.096 & 0.063 & 0.111 \\
\bottomrule
\end{tabular}}
\end{table}

\smallskip
A second type of application is illustrated by Table~\ref{tabOutliers}, which refers to a simulation experiment with $10^5$ replications per scenario. For simulated \iid\ $\Exp(1)$-samples of sizes $ n\in \{10,25,50,100\} $, about 10\,\% of the observations were randomly selected and contaminated by an additive outlier, namely by adding~5 to the selected observations. Note that the topic of outliers in exponential data received considerable interest in the literature \citep[pp.~528--530]{Johnson95}. Then, different estimators~$\hat{\lambda}_{f,\Exp}$ are computed from the contaminated data, where the first three choices of the weighting function~$f$ are characterized by a sublinear increase, whereas the fourth function, $f(x)=x$, corresponds to the default estimator $\hat{\lambda} = 1/\overline{X}$. Table~\ref{tabOutliers} shows that all MM estimators are affected by the outliers, \eg in terms of the strong negative bias. But comparing the four columns of bias and MSE values, respectively, it gets clear that the novel Stein-MM estimators are more robust against the outliers, having both lower bias and MSE than~$\hat{\lambda}$. Especially the choice $f(x)=\ln(1+x)$, a logarithmic weighting scheme, leads to a rather robust estimator. The relatively good performance of the Stein-MM estimators can be explained by the fact that the weighting functions increase sublinearly (which is also beneficial for bias reduction in non-contaminated data, recall the above discussion), so the effect of large observations is damped. To sum up, by choosing an appropriate weighting function~$f$ within the Stein-MM estimator~$\hat{\lambda}_{f,\Exp}$, one cannot only achieve a reduced bias and MSE, but also a reduced sensitivity towards outlying observations.

\section{Stein Estimation of Inverse Gaussian Distribution}
\label{Stein Estimation of Inverse-Gaussian Distribution}
%
Like the exponential distribution considered in the previous Section~\ref{Stein Estimation of Exponential Distribution}, the IG-distribution with parameters $\mu,\lambda>0$, abbreviated as $\IG(\mu,\lambda)$, has positive support, where the pdf is given by
\begin{align*}
	\phi(x)\ =\ \sqrt{\frac{\lambda}{2\pi}}\, e^{\lambda/\mu}\, x^{-3/2}\,\exp\!\bigg(-\frac{\lambda}{2\mu}\Big(\frac{x}{\mu}+\frac{\mu}{x}\Big)\bigg)\quad\text{for } x>0, \text{ and 0 otherwise}.
\end{align*}
The IG-distribution is commonly used as a lifetime model (as it can be related to the first-passage time in random walks), but it may also simply serve as a distribution with positive skewness and, thus, as an alternative to, \eg the lognormal distribution \citep[see][]{FolksCh78}. Detailed surveys about the properties and applications of $\IG(\mu,\lambda)$, and on many further references, can be found in \citet{FolksCh78,sesh99} as well as in \citet[Chapter 15]{Johnson95}. In what follows, the moment properties of $X\sim\IG(\mu,\lambda)$ are particularly relevant. We have $ \e[X]=\mu $, $ \e[1/X]=1/\mu +1/\lambda$, and $ \V[X] =\mu^3/\lambda$. In particular, positive and negative moments are related to each other by
\begin{align}\label{MomIG}
	\e\Big[ X^{-k} \Big]=\frac{\e[X^{k+1}]}{\mu^{2k+1}}
	\quad\Leftrightarrow\quad
	\e\Big[ (\mu/X)^{k} \Big]=\e\Big[(X/\mu)^{k+1}\Big]\quad\text{for } k\in\bbn_0,
\end{align}
see \citet[p.~372]{Tweed57a} as well as the aforementioned surveys.

\begin{bem}
\label{bemExpGoFtest}
At this point, let us briefly recall the discussion in Section~\ref{Stein Estimation of Exponential Distribution} (p.~\pageref{GoF-test}), where we highlighted the property that for \iid\ exponential samples, the quotient $\overline{f'(X)} / \overline{f(X)}$ has an (asymptotically) unique mean for different~$f$. From counterexamples, it is easily seen that this property is not true for $\IG(\mu,\lambda)$-data. The Delta method implies that the mean of $\overline{f'(X)} / \overline{f(X)}$ is asymptotically equal to $\e[f'(X)]\big/\e[f(X)]$, which equals
\begin{itemize}
	\item $1/\e[X] = 1/\mu$ for $f(x)=x$, but
	\item $2\e[X]/\e[X^2] = 2\mu\big/\big(\mu^2(1+\tfrac{\mu}{\lambda})\big) = 2\lambda\big/\big(\mu(\lambda+\mu)\big)$ for $f(x)=x^2$.
\end{itemize}
\end{bem}
From now on, let $ X_1,\ldots,X_n $ be an \iid\ sample from $\IG(\mu,\lambda) $, which shall be used for parameter estimation. Here, one obviously estimates~$\mu$ by the sample mean~$\overline{X}$, but the estimation of~$\lambda$ is more demanding. In the literature, the MM and ML estimation of~$\lambda$ have been discussed (see the details below), while our aim is to derive a generalized MM estimator with improved bias and MSE properties based on a Stein identity. In fact, as we shall see, our proposed approach can be understood as a unifying framework that covers the ordinary MM and ML estimator as special cases. A Stein identity for $\IG(\mu,\lambda)$ has been derived by \citet[p.~172]{Koudou14}, which states that
\begin{align}\label{Messi}
	X\sim\IG(\mu,\lambda) \quad \text{iff} \quad \e\big[f(X)\,(\lambda X^2-\mu^2 X-\lambda\mu^2)\big]\ =\ 2\mu^2\ \e\big[X^2f'(X)\big]
\end{align}
holds for all differentiable functions $f: (0;\infty)\to\bbr$ with $  \lim_{x \rightarrow 0}  f(x)\,\phi(x) =\lim_{x \rightarrow \infty}  f(x)\,\phi(x)=0 $. Solving \eqref{Messi} in $\lambda$ and using the sample moments $ \overline{h(X)}=\tfrac{1}{n}\sum_{i=1}^{n} h(X_i)$ instead of $\e\big[h(X)\big]$ (where~$h$ might be any of the functions involved in \eqref{Messi}), the class of Stein-MM estimators for $ \lambda $ is obtained as
\begin{align}\label{IGSteinMM}
	\hat{\lambda}_{f,\IG}
	=\frac{\, \overline{X}^2\big( 2\, \overline{X^2f'(X)}+\overline{Xf(X)}\, \big) \,}{\overline{X^2f(X)}-\overline{X}^2\, \overline{f(X)}}.
\end{align}
Here, the ordinary MM estimator of $ \lambda>0 $, \ie $ \hat{\lambda}_{\textup{MM}}=\overline{X}^3/S^2 $ with~$S^2$ denoting the empirical variance \citep{Tweed57b}, is included as the special case $ f\equiv 1 $, whereas the ML estimator $\hat{\lambda}_{\textup{ML}}=\overline{X}\big/\big(\overline{X}\cdot\overline{1/X}-1 \big)$ \citep{Tweed57a} follows for $ f(x)=1/x $. Hence, \eqref{IGSteinMM} provides a unifying estimation approach that covers the established estimators as special cases. 

\begin{bem}
\label{bemIG_Ebner}
At this point, a reference to Example~2.9 in \citet{ebner23} is necessary. As already mentioned in Section~\ref{Introduction}, also \citet{ebner23} proposed a Stein-MM estimator for the IG-distribution, which, however, differs from the one developed here. The crucial difference is given by the fact that \citet{ebner23} tried a joint estimation of~$(\mu,\lambda)$ based on \eqref{Messi}, namely by jointly solving two equations that are implied by \eqref{Messi} if using two different weight functions~$f_1\not= f_2$. The resulting class of estimators, however, does not cover the existing MM and ML estimators, so \citet{ebner23} did not pursue the Stein-MM estimation of the IG-distribution further. By contrast, as we did not see notable potential for improving the estimation of~$\mu$ by~$\overline{X}$ (recall the diverse optimality properties of the sample mean as an estimator of the population mean \citep[e.\,g.,][]{shuster82}), we used \eqref{Messi} to only derive an estimator for~$\lambda$. In this way, we were able to recover both the MM and ML estimator of~$\lambda$ within \eqref{IGSteinMM}. 
\end{bem}
For deriving the asymptotic distribution of our general Stein-MM estimator~$\hat{\lambda}_{f,\IG}$ from \eqref{IGSteinMM}, we first define the vectors $\Z_i $ with $ i=1,\ldots,n $ as 
\begin{align}\label{IGZi}
	\Z_i\coloneqq\Big(X_i,\, f(X_i),\, X_if(X_i),\, X_i^2f(X_i),\, X_i^2f'(X_i)\Big)^\top.
\end{align}
Their mean equals
\begin{align}\label{IGZiMean}
	\bmu_Z\coloneqq\e[\Z_i]=\Big(\mu,\, \mu_f(0,1,0),\, 
	\mu_f(1,1,0),\, \mu_f(2,1,0),\, \mu_f(2,0,1)\Big)^\top.
\end{align}
Then, the following CLT holds.

\begin{satz}\label{CLT_IG}
	If $X_1,\ldots,X_n$ are \iid\ according to $ \IG(\mu,\lambda) $, then the sample mean~$\overline{\Z}$ of $\Z_1,\ldots,\Z_n$ according to \eqref{IGZi} is asymptotically normally distributed as
	\begin{align*}
		\sqrt{n}\big(\overline{\Z}-\bmu_Z\big) \ \xrightarrow{\text{d}}\ \norm\big(\mathbf{0}, \bSigma\big)\quad \text{with}\quad \bSigma=(\sigma_{ij})_{i,j=1,\ldots,5},
	\end{align*}
	where $ \norm(\mathbf{0}, \bSigma) $ denotes the multivariate normal distribution, and where the covariances are given as 
	\begin{align*}
		&\sigma_{11}=\mu^3/\lambda, && \sigma_{23}= \mu_f(1,2,0)-\mu_f(0,1,0)\cdot\mu_f(1,1,0),\\
		&\sigma_{12}=\mu_f(1,1,0)-\mu\cdot\mu_f(0,1,0), && \sigma_{24}=\mu_f(2,2,0)-\mu_f(0,1,0)\cdot\mu_f(2,1,0) ,\\
		&\sigma_{13}=\mu_f(2,1,0)-\mu\cdot\mu_f(1,1,0), && \sigma_{25}=\mu_f(2,1,1)-\mu_f(0,1,0)\cdot\mu_f(2,0,1) ,\\
		&\sigma_{14}=\mu_f(3,1,0)-\mu\cdot\mu_f(2,1,0), && \sigma_{34}=\mu_f(3,2,0) -\mu_f(1,1,0)\cdot\mu_f(2,1,0),\\
		&\sigma_{15}=\mu_f(3,0,1)-\mu\cdot\mu_f(2,0,1), && \sigma_{35}= \mu_f(3,1,1)-\mu_f(1,1,0)\cdot\mu_f(2,0,1),\\
		&\sigma_{22}=\mu_f(0,2,0)-\mu_f^2(0,1,0), && \sigma_{45}= \mu_f(4,1,1)-\mu_f(2,1,0)\cdot\mu_f(2,0,1),\\
		&\sigma_{33}=\mu_f(2,2,0)-\mu_f^2(1,1,0), && \sigma_{55}= \mu_f(4,0,2)-\mu_f^2(2,0,1),\\
		&\sigma_{44}=\mu_f(4,2,0)-\mu_f^2(2,1,0).
	\end{align*}
\end{satz}
The proof of Theorem~\ref{CLT_IG} is provided by Appendix~\ref{Proof of Theorem CLT_IG}.

\smallskip
In the second step of deriving the asymptotics of~$\hat{\lambda}_{f,\IG}$, we define the function $ g(x_1,x_2,x_3,x_4,x_5)\coloneqq x_1^2(2x_5+x_3)/(x_4-x_1^2x_2)$. Then, $ \hat{\lambda}_{f,\IG}=g(\overline{\Z}) $ and $ \lambda=g(\bmu_Z) $. Applying the Delta method \citep{serfling80} to Theorem~\ref{CLT_IG}, the following result follows.

\begin{satz}\label{Pele}
	Let $X_1,\ldots,X_n$ be \iid\ according to $ \IG(\mu,\lambda) $, and define $ \vartheta_{f,\IG}\coloneqq  \mu_f(2,1,0)-\mu^2\mu_f(0,1,0)$. Then,  $ \hat{\lambda}_{f,\IG}$ is asymptotically normally distributed, where the asymptotic variance and bias, respectively, are given by 
	{\small%
	\begin{align*}
		\sigma_{f,\IG}^2&=\frac{1}{n}\Bigg[\, \frac{\mu^4}{\vartheta_f^2}\Bigg[\mu_f(2,2,0)-\mu_f(1,1,0)\Big(\frac{\lambda \vartheta_f}{\mu ^2}+2\mu_f(2,0,1)\Big) +4\mu_f(3,1,1)+4\mu_f(4,0,2)
		\\[1ex]&\quad-4\mu_f^2(2,0,1)\Bigg]+\frac{2\lambda\mu}{\vartheta_f^2}\Bigg[\mu^3\Bigg(\mu_f(1,2,0)+2\mu_f(2,1,1)-\frac{\lambda \vartheta_f}{\mu ^2}\cdot\mu_f(0,1,0)\Bigg)
		\\[1ex]&\quad-\mu\Big(\mu_f(3,2,0)+2\mu_f(4,1,1)\Big)+\mu_f(2,1,0)\Big(4\mu_f(2,1,0)+4\mu_f(3,0,1)-\frac{\lambda \vartheta_f}{\mu}\Big)\Bigg]
		\\[1ex]&\quad+\frac{\lambda^2}{\mu\vartheta_f^2}\Bigg[ \mu^5\cdot\mu_f(0,2,0)+\mu^3\cdot\mu_f(0,1,0)\Big(\vartheta_f+\mu_f(2,1,0)\Big) -2\mu^3\cdot\mu_f(2,2,0)
		\\[1ex]&\quad+4\mu_f(2,1,0)\Big(\mu^2\cdot\mu_f(1,1,0)-\mu^3\cdot\mu_f(0,1,0)-\mu_f(3,1,0) \Big)
		\\[1ex]&\quad+\mu\Big(3\mu_f^2(2,1,0)+\mu_f(4,2,0)\Big)\Bigg]\, \Bigg],
	\end{align*}}
	and
	{\small%
	\begin{align*}
		\mathbb{B}_{f,\IG}&=\frac{1}{n}\Bigg[\frac{1}{\mu\vartheta_f^2}\Bigg[\mu^2\cdot\mu_f(2,1,0)\Big(\mu_f(2,1,0)+3\mu^2\cdot\mu_f(0,1,0)\Big)
		+\lambda\bigg(\mu^5\cdot\mu_f(0,2,0) \\[1ex]&\quad+\mu\cdot\mu_f(4,2,0)-2 \mu_f(3,1,0)\cdot\mu_f(2,1,0)+4\mu^2\cdot\mu_f(1,1,0)\cdot\mu_f(2,1,0)
		\\[1ex]&\quad-2\mu^3\cdot\mu_f(2,2,0)\bigg)\Bigg]+\frac{\mu}{\vartheta_f^3}\Bigg[
		\mu_f(2,1,0)\Bigg( 2\mu_f^2(2,1,0)
		-\mu\cdot\mu_f(3,2,0)
		\\[1ex]&\quad+4\mu_f(2,1,0)\cdot\mu_f(3,0,1)-2\mu\cdot\mu_f(4,1,1)+\mu^3\Big(\mu_f(1,2,0)+2\mu_f(2,1,1)\Big)\Bigg)
		\\[1ex]&\quad-\mu^2\cdot\mu_f(0,1,0)\Bigg(2\mu_f(3,1,0)\Big( 2\mu_f(2,0,1)+\mu_f(1,1,0)\Big)
		\\[1ex]&\quad+2\mu_f^2(2,1,0)+4\mu_f(2,1,0)\cdot\mu_f(3,0,1) +\mu^3\Big(\mu_f(1,2,0)+2\mu_f(2,1,1)\Big)
		\\[1ex]&\quad-\mu\Big(\mu_f(3,2,0)+2\mu_f(4,1,1) \Big)\Bigg)
		\Bigg]\Bigg].
	\end{align*}}
\end{satz}
The proof of Theorem~\ref{Pele} is provided by Appendix~\ref{Proof of Theorem Pele}. 

\smallskip
Before we discuss the effect of~$f$ on bias and MSE of~$\hat{\lambda}_{f,\IG}$, let us first consider the special cases of the ordinary MM and ML estimator. Their asymptotics are immediate consequences of Theorem~\ref{Pele}. For the MM estimator $\hat{\lambda}_{\textup{MM}}$, we have to choose $f\equiv 1$ such that $f'\equiv 0$. As a consequence,
\ba\label{mu1klm}
	\mu_1(k,l,m)= \e[X^k]\quad\text{if }\quad m=0, \quad\text{and 0 otherwise}.
\ea
This leads to a considerable simplification of Theorem~\ref{Pele}, see Appendix~\ref{Proof_korIG_MM}, which is summarized in the following corollary.

\begin{kor}\label{korIG_MM}
	Let $X_1,\ldots,X_n$ be \iid\ according to $ \IG(\mu,\lambda) $, then $ \hat{\lambda}_{\textup{MM}}=\overline{X}^3/S^2 $ is asymptotically normally distributed with asymptotic variance $\sigma_{\textup{MM}}^2 = \frac{2}{n}\,\lambda(\lambda+3\mu)$ and bias $\mathbb{B}_{\textup{MM}} = \frac{3}{n}\,(\lambda+3\mu)$.
\end{kor}
While we are not aware of a reference providing these asymptotics, they can be verified by using \citet[p.~704]{Tweed57b}. There, normal asymptotics for the reciprocal $1/\hat{\lambda}_{\textup{MM}}$ are provided: $\sqrt{n}\big(\hat{\lambda}_{\textup{MM}}^{-1}-\lambda^{-1}\big)\ \sim\norm\big(0,\ 2(1+3\mu/\lambda)/\lambda^2\big)$. Applying the Delta method with $g(x)=1/x$ and $g'(x)=-1/x^2$ to it, we conclude that $\sqrt{n}(\hat{\lambda}_{\textup{MM}}-\lambda)$ has the asymptotic variance $\lambda^4\cdot 2(1+3\mu/\lambda)/\lambda^2 = 2\lambda\, (\lambda+3\mu)$ like in Corollary~\ref{korIG_MM}.

\smallskip
Next, we consider the special case of the ML estimator $\hat{\lambda}_{\textup{ML}}$, which follows by choosing $f(x)= 1/x$ such that $f'(x)=-1/x^2$. Again, the joint moments $\mu_f(k,l,m)$ simplify a lot:
\ba\label{mu1Xklm}
	\mu_{1/x}(k,l,m) = (-1)^m\,\e[X^{k-l-2m}]  \quad\text{for all } k,l,m\in\bbn_0.
\ea
Together with Theorem~\ref{Pele}, see Appendix~\ref{Proof_korIG_ML}, we get the following corollary.

\begin{kor}\label{korIG_ML}
	Let $X_1,\ldots,X_n$ be \iid\ according to $ \IG(\mu,\lambda) $, then $\hat{\lambda}_{\textup{ML}}=\overline{X}\big/\big(\overline{X}\cdot\overline{1/X}-1 \big)$ is asymptotically normally distributed with asymptotic variance $\sigma_{\textup{ML}}^2 = \frac{2}{n}\,\lambda^2$ and bias $\mathbb{B}_{\textup{ML}} = \frac{3}{n}\,\lambda$.
\end{kor}
Comparing Corollaries~\ref{korIG_MM} and~\ref{korIG_ML}, it is interesting to note that the MM~estimator has larger asymptotic bias and variance than the ML~estimator: $\sigma_{\textup{MM}}^2 = \sigma_{\textup{ML}}^2+\tfrac{6\lambda\mu}{n}$ and $\mathbb{B}_{\textup{MM}} =\mathbb{B}_{\textup{ML}}+\tfrac{9\mu}{n}$.
To verify the asymptotics of Corollary~\ref{korIG_ML}, note that the ML~estimator~$\hat{\lambda}_{\textup{ML}}$ has been shown to follow an inverted-$\chi^2$ distribution: $\hat{\lambda}_{\textup{ML}}\ \sim\ n\,\lambda\cdot \text{Inv-}\chi^2_{n-1}$ \citep[see][p.~368]{Tweed57a}. Using the formulae for mean and variance of~$\text{Inv-}\chi^2_{n-1}$ \citep[see][p.~431]{bernardo94}, we get
$$
\e[\hat{\lambda}_{\textup{ML}}] = \tfrac{n}{n-3}\,\lambda = (1+\tfrac{3}{n-3})\,\lambda \approx (1+\tfrac{3}{n})\,\lambda,
\quad
\V[\hat{\lambda}_{\textup{ML}}]
 = \tfrac{2\,n^2}{(n-3)^2(n-5)}\,\lambda^2
\approx \tfrac{2}{n}\,\lambda^2
$$
for large~$n$, which agrees with Corollary~\ref{korIG_ML}. 

\begin{table}[!ht]
	\centering
	\caption{Simulated vs.\ asymptotic mean and standard deviation of estimator $\hat{\lambda}_{f,\IG}$ from \eqref{IGSteinMM}.}
	\label{tabIGsimulation}
	
	\smallskip
	\resizebox{\linewidth}{!}{
		\begin{tabular}{lr|ll|ll@{\qquad\quad}lr|ll|ll}
			\toprule
			&& \multicolumn{2}{c|}{Mean} & \multicolumn{2}{c}{Std.\,dev.} &&& \multicolumn{2}{c|}{Mean} & \multicolumn{2}{c}{Std.\,dev.} \\
			$f(x)$ & $n$ & sim & asym & sim & asym. & $f(x)$ & $n$ & sim & asym & sim & asym. \\ \midrule
			\multicolumn{6}{l}{$(\mu,\lambda)=(1,1)$} & \multicolumn{6}{l}{$(\mu,\lambda)=(3,1)$} \\
			\midrule
			1 & 100 & 1.103 & 1.120 & 0.258 & 0.283 & 1 & 100 &  1.213	&1.300	&0.365&	0.447   \\			
			& 250 & 1.045 &1.048  &0.169  & 0.179 &  & 250 &  1.098	&1.120&	0.245&	0.283  \\
			& 500 & 1.023 & 1.024 & 0.122 & 0.126 &  & 500 & 1.053	&1.060&	0.180&	0.200  \\
			\midrule
			$x^{-1/3}$ & 100 &1.125	&1.138	&0.306&	0.312& $x^{-1/3}$ & 100 &1.243&	1.294&	0.442&	0.451 \\
			& 250 &1.053	&1.055	&0.193	&0.198&& 250 &1.104	&1.118	&0.274	&0.285 \\
			& 500 &1.027	&1.028	&0.138	&0.140&&   500 &1.055	&1.059	&0.194	&0.202 \\
			\midrule
			$x^{-2/3}$ & 100 &1.026	&1.025	&0.158	&0.151& $x^{-2/3}$ & 100 &1.020&	1.019	&0.160	&0.156\\
			& 250 &1.010	&1.010	&0.098	&0.096&  & 250 &1.008	&1.008	&0.100	&0.099 \\
			& 500 &1.006	&1.005	&0.069	&0.068&&   500 &1.004	&1.004	&0.070	&0.070\\
			\midrule
			$1/x$ & 100 &1.031	&1.030	&0.149	&0.141&  $1/x$ &100&1.031	&1.030&	0.149&	0.141\\
			& 250 &1.012	&1.012	&0.091	&0.089 && 250 &1.012	&1.012	&0.091	&0.089 \\
			& 500 &1.006	&1.006	&0.064	&0.063&& 500 &1.006	&1.006	&0.064	&0.063 \\
			\midrule
			$x^{-4/3}$ & 100 &1.032	&1.031	&0.151	&0.143& $x^{-4/3}$ & 100 &1.033&	1.032	&0.154	&0.146\\
			& 250 &1.013	&1.013	&0.093	&0.091&& 250 &1.013	&1.013	&0.094	&0.093\\
			& 500 &1.007	&1.006	&0.065	&0.064&& 500 &1.007	&1.006	&0.066	&0.065 \\
			\midrule
			\multicolumn{6}{l}{$(\mu,\lambda)=(1,3)$} & \multicolumn{6}{l}{$(\mu,\lambda)=(3,3)$} \\
			\midrule
			1 & 100 &3.172	&3.180&	0.595&	0.600& 1 & 100 &3.309	&3.360	&0.775&	0.849 \\
			& 250 &3.071&	3.072&	0.378&	0.379 & &250 &3.134	&3.144	&0.506	&0.537 \\
			& 500 &3.036&	3.036&	0.267&	0.268 & & 500 &3.069&	3.072&	0.366&	0.379 \\
			\midrule
			$x^{-1/3}$ & 100 &3.207	&3.216	&0.677	&0.675& $x^{-1/3}$ & 100 &3.374	&3.415	&0.918	&0.937\\
			& 250 &3.085	&3.086	&0.427	&0.427&& 250 &3.159	&3.166	&0.580	&0.593 \\
			& 500 &3.043	&3.043	&0.301	&0.302&& 500 &3.081	&3.083	&0.413	&0.419 \\
			\midrule
			$x^{-2/3}$ & 100 &3.087	&3.085	&0.462	&0.440&  $x^{-2/3}$ & 100 &3.078&	3.076	&0.475	&0.454\\
			& 250 &3.035	&3.034	&0.284	&0.278&& 250 &3.031	&3.031	&0.293	&0.287 \\
			& 500 &3.018	&3.017	&0.199	&0.197&& 500 &3.017	&3.015	&0.206	&0.203 \\
			\midrule
			$1/x$ & 100 &3.093	&3.090	&0.448	&0.424&  $1/x$ & 100 &3.093	&3.090&	0.448&	0.424\\
			& 250 &3.037	&3.036	&0.274	&0.268&& 250 &3.037	&3.036	&0.274	&0.268 \\
			& 500 &3.019	&3.018	&0.192	&0.190&& 500 &3.019	&3.018	&0.192	&0.190 \\
			\midrule
			$x^{-4/3}$ & 100 &3.095	&3.092	&0.449	&0.426&  $x^{-4/3}$ & 100 &3.097&	3.094	&0.453	&0.430 \\
			& 250 &3.038	&3.037	&0.275	&0.269&& 250 &3.039	&3.038	&0.278	&0.272 \\
			& 500 &3.020	&3.018	&0.193	&0.191&& 500 &3.020	&3.019	&0.195	&0.192 \\
			\bottomrule
	\end{tabular}}
\end{table}

\begin{figure}[t]
	\center\footnotesize
	(a)\hspace{-3ex}\includegraphics[scale=0.5, viewport=0 45 335 305, clip=]{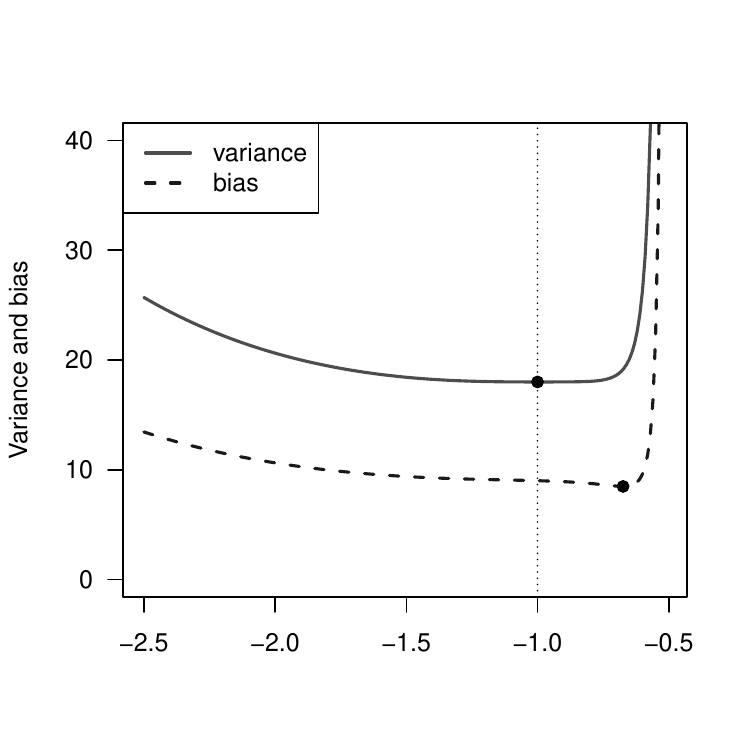}~$a$
	\qquad
	(b)\hspace{-3ex}\includegraphics[scale=0.5, viewport=0 45 335 305, clip=]{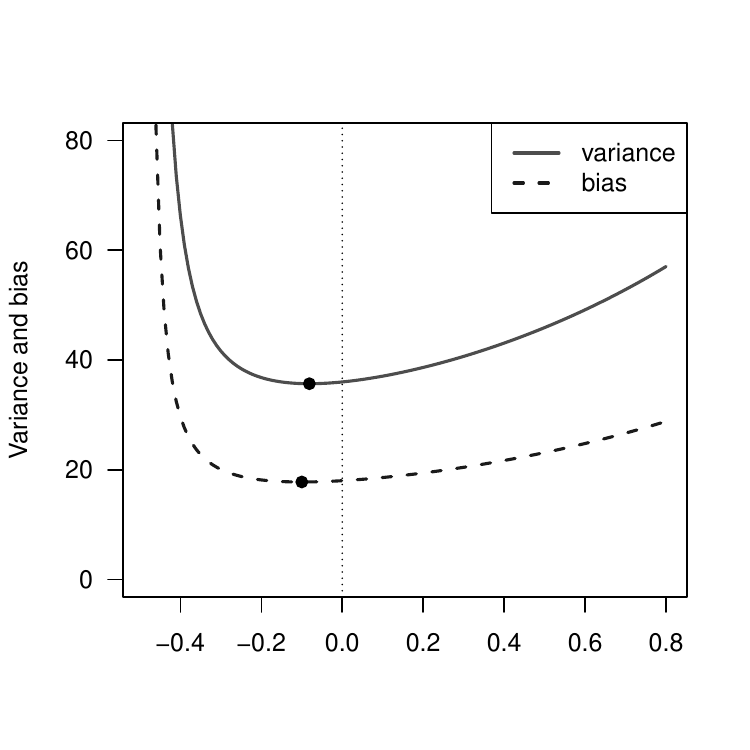}$a$
\\[2ex]
	(c)\hspace{-3ex}\includegraphics[scale=0.5, viewport=0 45 335 305, clip=]{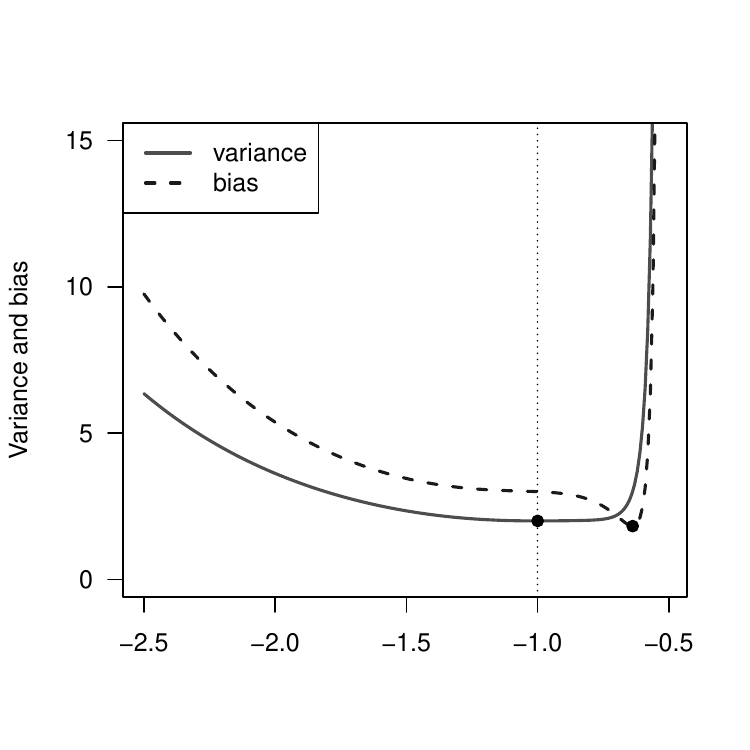}~$a$
	\qquad
	(d)\hspace{-3ex}\includegraphics[scale=0.5, viewport=0 45 335 305, clip=]{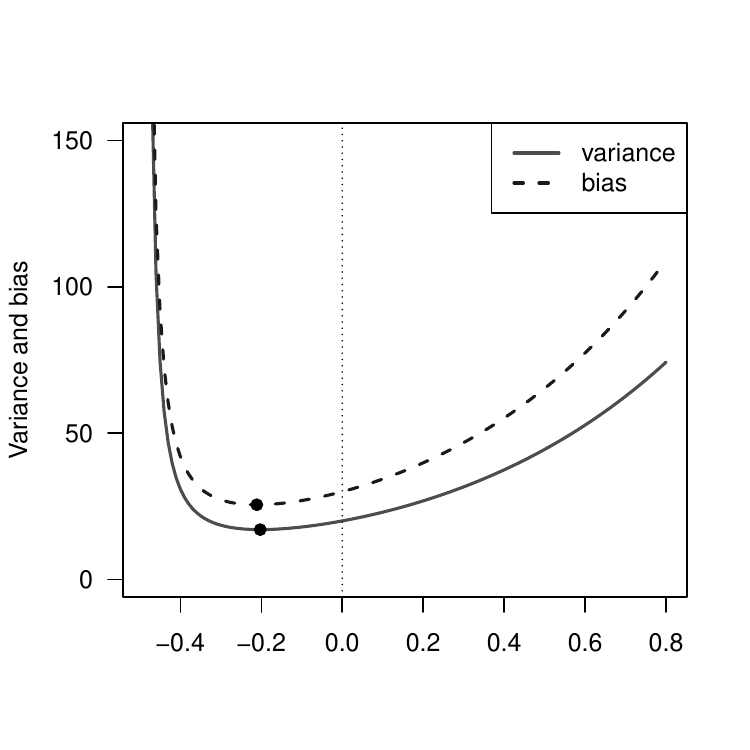}$a$
	\caption{Plots of $n\,\sigma^2 _{f_a,\IG} $ and $ n\,\mathbb{B}_{f_a,\IG} $, where points indicate minimal variance and bias values. Scenarios $(\mu,\lambda)=(1,3)$ with (a) $ a\in (-2.5;-0.5) $ and (b) $ a\in (-0.5;0.8) $, and $(\mu,\lambda)=(3,1)$ with (c) $ a\in (-2.5;-0.5) $ and (d) $ a\in (-0.5;0.8) $. Dotted lines at $a=-1$ and $a=0$ correspond to default ML and MM estimator, respectively.}
	\label{fig_VarBias_IG_xa}
\end{figure}

\begin{figure}[t]
	\center\footnotesize
	(a)\hspace{-3ex}\includegraphics[scale=0.5, viewport=0 45 335 305, clip=]{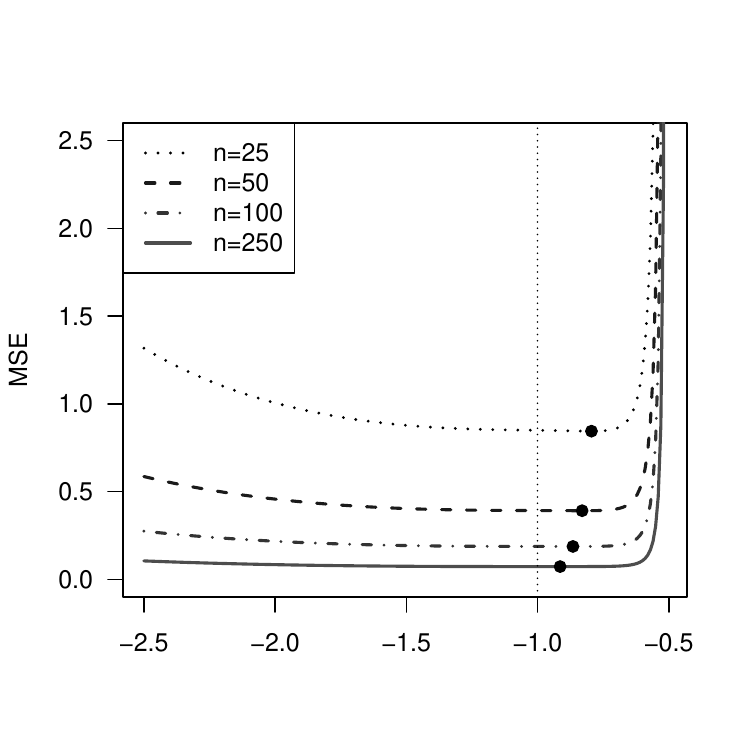}~$a$
	\qquad
	(b)\hspace{-3ex}\includegraphics[scale=0.5, viewport=0 45 335 305, clip=]{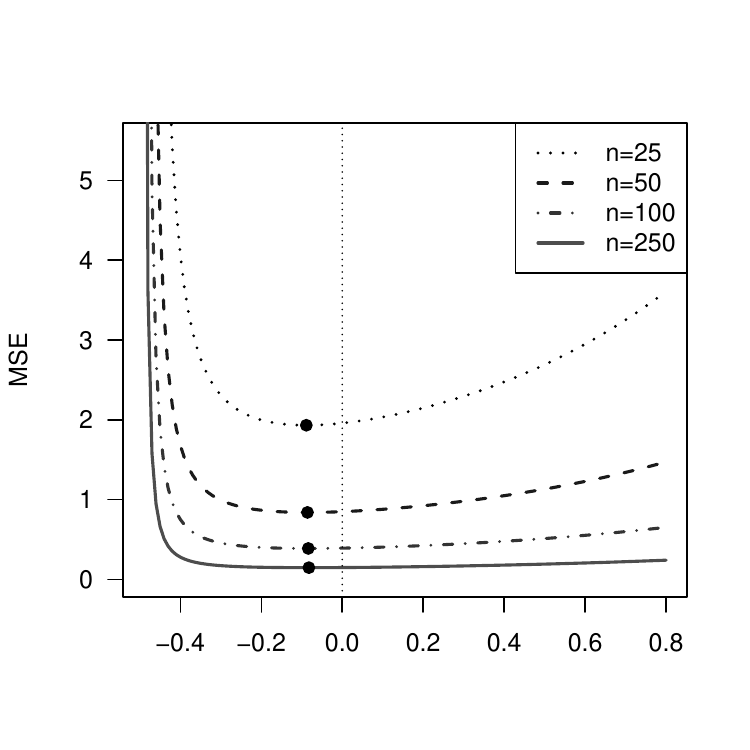}$a$
\\[2ex]
	(c)\hspace{-3ex}\includegraphics[scale=0.5, viewport=0 45 335 305, clip=]{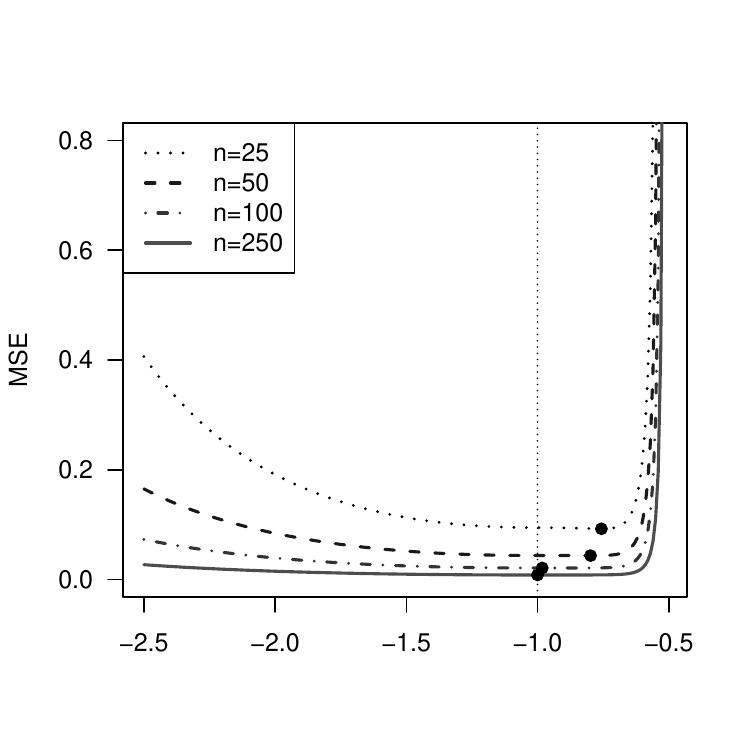}~$a$
	\qquad
	(d)\hspace{-3ex}\includegraphics[scale=0.5, viewport=0 45 335 305, clip=]{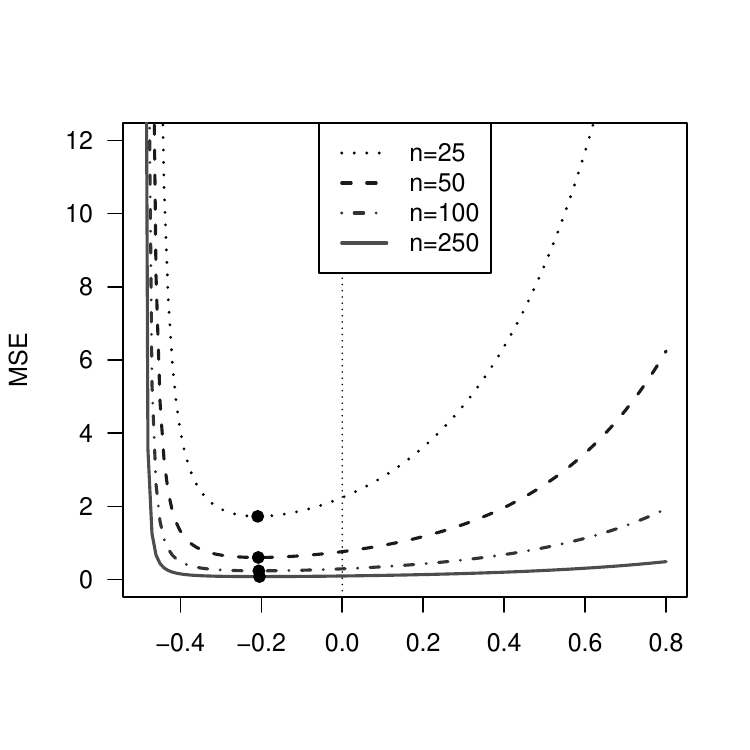}$a$
	\caption{Plots of MSE$ _{f_a,\IG} $, where points indicate minimal MSE values. Scenarios $(\mu,\lambda)=(1,3)$ with (a) $ a\in (-2.5;-0.5) $ and (b) $ a\in (-0.5;0.8) $, and $(\mu,\lambda)=(3,1)$ with (c) $ a\in (-2.5;-0.5) $ and (d) $ a\in (-0.5;0.8) $. Dotted lines at $a=-1$ and $a=0$ correspond to default ML and MM estimator, respectively.}
	\label{fig_MSE_IG_xa}
\end{figure}

\begin{bem}
\label{bemIGsimulation}
To analyze the performance of the asymptotics provided by Theorem~\ref{Pele} (and that of the special cases discussed in Corollaries~\ref{korIG_MM} and~\ref{korIG_ML}), if used as approximations to the true distribution of~$\hat{\lambda}_{f,\IG}$ for finite sample size~$n$, we did a simulation experiment with $10^5$ replications. The obtained results for various choices of~$(\mu,\lambda)$ and~$f(x)$ are summarized in Table~\ref{tabIGsimulation}. It can be recognized that the asymptotic approximations for mean and standard deviation generally agree quite well with their simulated counterparts. Only for the case $f(x)\equiv 1$ (the default MM~estimator) and sample size $n=100$, we sometimes observe stronger deviations. But in the large majority of estimation scenarios, we have a close agreement such that the conclusions derived from the asymptotic expressions are meaningful for finite sample sizes as well. 
\end{bem}
In analogy to our discussion in Section~\ref{Stein Estimation of Exponential Distribution}, let us now analyze the performance of the Stein-MM estimator $\hat{\lambda}_{f,\IG}$ for the weight functions $f_a: (0;\infty)\to (0;\infty)$, $f_a(x)=x^a$ with $a\in\bbr\setminus\{-\frac{1}{2}\}$. Recall that this class of weight functions cover the default MM estimator $\hat{\lambda}_{\textup{MM}}$ for $a=0$ and the ML estimator $\hat{\lambda}_{\textup{ML}}$ for $a=-1$. The choice $a=-\frac{1}{2}$ (right in the middle between these two special cases) has to be excluded as it leads to a degenerate estimator $\hat{\lambda}_{f_a,\IG}$ according to \eqref{IGSteinMM}. For this reason, the subsequent analyses in Figures~\ref{fig_VarBias_IG_xa} and~\ref{fig_MSE_IG_xa} are done separately for $a<-\frac{1}{2}$ (plots on left-hand side, covering the ML~estimator) and $a>-\frac{1}{2}$ (plots on right-hand side, covering the MM~estimator). 

\smallskip
Let us start with the analysis of asymptotic bias and variance in Figure~\ref{fig_VarBias_IG_xa}. The upper and lower panel consider two different example situations, namely $(\mu,\lambda)=(1,3)$ and $(\mu,\lambda)=(3,1)$, respectively, while left-hand and right-hand side are separated by the pole at $a=-\frac{1}{2}$. The right-hand side shows that the default MM estimator is neither (locally) optimal in terms of asymptotic bias nor in terms of variance. In fact, the optimal~$a$ for $a>-\frac{1}{2}$ is around~$-0.1$ for $(\mu,\lambda)=(1,3)$, and around~$-0.2$ for $(\mu,\lambda)=(3,1)$. However, comparing the actual values at the Y-axis to those of the plots on the left-hand side, we recognize that the asymptotic bias and variance get considerably smaller for some region with $a<-\frac{1}{2}$. In particular, the ML~estimator is clearly superior to the MM estimator, and as shown by Figures~\ref{fig_VarBias_IG_xa}\,(a) and~(c), the ML estimator is even optimal in terms of the asymptotic variance. It has to be noted, however, that the curve corresponding to the asymptotic variance is rather flat around $a=-1$, so moderate deviations from $a=-1$ do not have a notable effect on the variance. Thus, it is important to also consider the optimum bias, which is reached for some~$a$ around~$-0.65$ in both~(a) and~(c). So it appears to be advisable to choose an $a>-1$ for optimal overall estimation performance.

\smallskip
This is confirmed by Figure~\ref{fig_MSE_IG_xa}, where the asymptotic MSE is shown for various sample sizes~$n$ and the same scenarios as in Figure~\ref{fig_VarBias_IG_xa}. While the ML~estimator approaches the MSE-optimum for increasing~$n$, we get an improved performance for smaller sample sizes if choosing $a\in (-1;-0.5)$ appropriately (\eg $a\approx -0.8$ if $n\leq 50$). Generally, an analogous recommendation holds for $a>-\frac{1}{2}$ in parts~(b) and~(d), with MSE-optima at~$a$ around~$-0.1$ and~$-0.2$, respectively, but much smaller MSE~values can be reached for $a<-\frac{1}{2}$. To sum up, the default MM~estimator (and more generally, Stein-MM estimators $\hat{\lambda}_{f_a,\IG}$ with $a>-\frac{1}{2}$) are not recommended for practice due to their rather large bias, variance, and thus MSE, while the ML~estimator constitutes at least a good initial choice for estimating~$\lambda$, being optimal in terms of asymptotic variance. However, unless the sample size~$n$ is very large, an improved MSE performance can be achieved by reducing~$a$ to an appropriate value in $(-1;-0.5)$ in the second step and by computing the corresponding  Stein-MM estimate $\hat{\lambda}_{f_a,\IG}$. 

\begin{table}[t]
		\centering
		\caption{Runoff data from Example~\ref{bspIGdata}: Stein-MM estimates $\hat{\lambda}_{f_a,\IG}$ for different choices of function $f_a(x)=x^a$.}
		\label{Runoff_data}

\smallskip
\resizebox{\linewidth}{!}{
$\displaystyle
\begin{array}{r|c@{\qquad}c@{\qquad}c|c@{\qquad}c@{\qquad}c@{\qquad}c}
				\toprule
				a & -1.5 & -1 & -0.668 & -0.125 & -0.109 & 0 & 0.5 \\ 
				\text{\footnotesize Notes} &  & \text{\footnotesize (i)} & \text{\footnotesize (ii)} & \text{\footnotesize (iii)} & \text{\footnotesize (iv)} & \text{\footnotesize (v)} &  \\
				\midrule
				\hat{\lambda}_{f_a,\IG} & 1.423 & 1.440 & 1.429 & 1.511 & 1.511 & 1.512 & 1.529 \\
\bottomrule
\multicolumn{4}{l}{\text{\footnotesize (i)\quad ML estimate, $\sigma^2 _{f_a,\IG}$-optimal}} &
\multicolumn{2}{l}{\text{\footnotesize (ii)\quad $\mathbb{B}_{f_a,\IG}$-optimal}} &
\multicolumn{2}{l}{\text{\footnotesize (v)\quad MM estimate}} \\
\multicolumn{4}{l}{\text{\footnotesize (iii)\quad $\mathbb{B}_{f_a,\IG}$-optimal given $a>-0.5$}} &
\multicolumn{4}{l}{\text{\footnotesize (iv)\quad $\sigma^2 _{f_a,\IG}$-optimal given $a>-0.5$}}
		\end{array}
$}
\end{table}

\begin{bsp}
\label{bspIGdata}
As an illustrative data example, let us consider the $n=25$ runoff amounts at Jug Bridge in Maryland \citep[see][p.~272]{FolksCh78}, which are ``very well described by the inverse Gaussian distribution''. 
The parameter~$\mu$ is estimated by the sample mean as~$\approx 0.803$, and using the ML~estimator $\hat{\lambda}_{f_{-1},\IG}$ as an initial estimator for~$\lambda$, we get the value $\approx 1.440$. As outlined before, this initial model fit might now be used for searching estimators with improved performance. Some examples (together with further estimates for comparative purposes) are summarized in Table \ref{Runoff_data}. The ML~estimator ($a=-1$) is also optimal in asymptotic variance, whereas the bias-optimal choice is obtained for a somewhat larger value of~$a$, namely $a\approx -0.668$. The corresponding estimate is slightly lower than the ML~estimate, similar to the value for $a=-1.5$, and can thus be seen as a fine-tuning of the initial estimate. By contrast, a notable change in the estimate happens if we turn to $a>-0.5$. The ``constrained-optimal'' choices (optimal given that $a>-0.5$) as well as the MM~estimate lead to nearly the same values (around~1.51) and are thus visibly larger than the actually preferable estimates for $a<-0.5$. Also their variance and bias are about 2.5~times larger than those of the estimates for $a<-0.5$.
\end{bsp}

%
\section{Stein Estimation of Negative-binomial Distribution}
\label{Stein Estimation of Negative-binomial Distribution}
%
While the previous sections (and also the research by \citet{ebner23}) solely focussed on continuous distributions, let us now turn to the case of discrete-valued random variables. Here, the most relevant type are count random variables~$X$, having a quantitative range contained in~$\bbn_0$. The probably most well-known distributions for counts are Poisson and binomial distributions, both depending on the (normalized) mean as their only model parameter. But as already discussed in Section~\ref{Stein Estimation of Inverse-Gaussian Distribution}, there is hardly any potential for finding a better estimator of the mean than the sample mean, so we do not further discuss these distributions. Instead, we focus on another popular count distribution, namely the NB-distribution with parameters $ \nu>0$ and $\pi\in(0;1) $, abbreviated as $ \nb(\nu,\pi) $. Such $X\sim \nb(\nu,\pi) $ has the range~$\bbn_0$, probability mass function (pmf) $\Pro(X=x) = \binom{\nu+x-1}{x}\, (1-\pi)^x\, \pi^\nu$, and mean $\mu= \e[X]=\frac{\nu(1-\pi)}{\pi} $. By contrast to the equidispersed Poisson distribution, its variance $\sigma^2 := \V[X]=\frac{\nu(1-\pi)}{\pi^2} $ is always larger than the mean (overdispersion), which is an important property for applications in practice. A detailed survey about the properties of and estimators for the $\nb(\nu,\pi)$-distribution can be found in \citet[Chapter 5]{Johnson05}. Instead of the original parametrization by~$(\nu,\pi)$, it is often advantageous to consider either~$(\mu,\nu)$ or~$(\mu,\pi)$, where~$\nu$ or~$\pi$, respectively, serve as an additional dispersion parameter once the mean~$\mu$ has been fixed. In case of the $(\mu,\nu)$-parametrization, it holds that $\pi=\tfrac{\nu}{\nu+\mu}$ and $ \V[X]=\tfrac{\nu+\mu}{\nu}\,\mu $, whereas we get $\nu=\tfrac{\pi\,\mu}{1-\pi}$ and $ \V[X]=\tfrac{1}{\pi}\,\mu $ for the $(\mu,\pi)$-parametrization. Besides the ease of interpretation, these parametrizations are advantageous in terms of parameter estimation. While MM~estimation is rather obvious, namely~$\mu$ by~$\overline{X}$ and $\nu_{\textup{MM}}=\overline{X}^2/(S^2-\overline{X})$, $\pi_{\textup{MM}}=\overline{X}/S^2$, ML~estimation is generally demanding as there does not exist a closed-form solution, see the discussion by \citet{Kemp87}, \ie numerical optimization is necessary. However, there is an important exception: 
the NB's ML~estimator of the mean~$\mu$ is given by~$\overline{X}$ \citep[p.~867]{Kemp87}, 
\ie $\overline{X}$ is both the~MM and ML~estimator with its known appealing performance. So it suffices to find an adequate estimator for~$\nu$ or~$\pi$, respectively, the ML estimators of which do not have a closed-form expression. 

\smallskip
These difficulties in estimating~$\nu$ or~$\pi$, respectively, serve as our motivation for deriving a generalized MM~estimator. For this purpose, we consider the NB's Stein identity according to \citet[Lemma 1]{Brown99}, which can be expressed as either
\begin{align}
	\label{NB1}
	&X\sim\nb(\nu,\, \tfrac{\nu}{\nu+\mu}) \quad \text{iff} \nonumber
	\\& \quad\nu\,\e\big[X\,f(X)-\mu f(X+1)\big]=\mu\,\e\big[X\,\big(f(X+1)-f(X)\big)\big],\\
	\text{or}\nonumber\\
	\label{NB2}
	&X\sim\nb(\tfrac{\pi\,\mu}{1-\pi},\, \pi) \quad \text{iff} \nonumber
	\\& \quad \pi\,\e\big[(X-\mu)\, f(X+1)\big]=\e\big[X\big(f(X+1)-f(X)\big)\big],
\end{align}
for any function $ f $ such that $ \e\big[\vert f(X)\vert\big] $, $ \e\big[\vert f(X+1)\vert\big] $ exist. Note that the discrete difference $\Delta f(x) := f(x+1)-f(x)$ in \eqref{NB1} and \eqref{NB2} plays a similar role as the continuous derivative $f'(x)$ in the previous Stein identities \eqref{Lionel} and \eqref{Messi}. 

\smallskip
Stein-MM estimators are now derived by solving \eqref{NB1} in~$\nu$ or \eqref{NB2} in~$\pi$, respectively, and by using again sample moments $ \overline{h(X)}=\tfrac{1}{n}\sum_{i=1}^{n} h(X_i)$ instead of the involved population moments $\e\big[h(X)\big]$ (with~$\mu$ being estimated by~$\overline{X}$). 
As a result, the (closed-form) classes of Stein-MM estimators for~$\nu$ and~$\pi$ are obtained as
\begin{align}
	\label{NBSteinMM}
	\hat{\nu}_{f,\nb}	= \frac{\,  \overline{X}\, \overline{X\,\Delta f(X)} \,}{\, \overline{X\,f(X)}-\overline{X}\, \overline{f(X+1)}\, },\qquad \hat{\pi}_{f,\nb}= \frac{\,  \overline{X\,\Delta f(X)} \,}{\, \overline{X\,f(X+1)}-\overline{X}\, \overline{f(X+1)}\, }.
\end{align}
Note that the choice $f(x)=x$ (hence $\Delta f(x)=1$) leads to the default MM estimators given above. The ML~estimators are not covered by \eqref{NBSteinMM} this time, because they do not have a closed-form expression at all. Note, however, that the so-called ``weighted-mean estimator'' for~$\nu$ in (2.6) of \citet{Kemp87}, which was motivated as a kind of approximate ML~estimator, is covered by \eqref{NBSteinMM}, namely by choosing $f_\alpha(x)=\alpha^x$ with $\alpha\in (0;1)$. It is also worth pointing to \citet{savani06}, who define an estimator of~$\nu$ based on the moment~$\overline{f(X)}$ for some specified~$f$; their approach, however, usually does not lead to a closed-form estimator.

\bigskip
For deriving the asymptotic distribution of the general Stein-MM estimator~$	\hat{\nu}_{f,\nb}$ or $ \hat{\pi}_{f,\nb} $, respectively, we first define the vectors $\Z_i $ with $ i=1,\ldots,n $ as 
\begin{align}
	\label{NbZi}
	\Z_i\coloneqq\Big(X_i,\ f(X_i+1),\ X_i\, f(X_i),\ X_i\, f(X_i+1)\Big)^\top.
\end{align}
Their mean equals
\begin{align}\label{NbZiMean}
	\bmu_Z\coloneqq\e[\Z_i]=\Big(\mu,\ \tilde{\mu}_f(0,0,1),\ 
	\tilde{\mu}_f(1,1,0),\ \tilde{\mu}_f(1,0,1)\Big)^\top,
\end{align}
where we define $ \tilde{\mu}_f(k,l,m)\coloneqq \e[X^k\cdot f(X)^l\cdot f(X+1)^m]$ for any $k,l,m\in\bbn_0$. Then, the following CLT holds.

\begin{satz}
	\label{CLT_Nb}
	If $X_1,\ldots,X_n$ are \iid\ according to a negative binomial distribution, then the sample mean~$\overline{\Z}$ of $\Z_1,\ldots,\Z_n$ according to \eqref{NbZi} is asymptotically normally distributed as
	\begin{align*}
		\sqrt{n}\big(\overline{\Z}-\bmu_Z\big) \ \xrightarrow{\text{d}}\ \norm\big(\mathbf{0}, \bSigma\big)\quad \text{with}\quad \bSigma=(\sigma_{ij})_{i,j=1,\ldots,4},
	\end{align*}
	where $ \norm(\mathbf{0}, \bSigma) $ denotes the multivariate normal distribution, and where the covariances are given as 
		\begin{align*}
		&\sigma_{11}=\sigma^2, && \sigma_{23}= \tilde{\mu}_f(1,1,1)-\tilde{\mu}_f(0,0,1)\cdot\tilde{\mu}_f(1,1,0),\\
		&\sigma_{12}=\tilde{\mu}_f(1,0,1)-\mu\cdot\tilde{\mu}_f(0,0,1), &&		 \sigma_{24}=\tilde{\mu}_f(1,0,2)-\tilde{\mu}_f(0,0,1)\cdot\tilde{\mu}_f(1,0,1) ,\\
		&\sigma_{13}=\tilde{\mu}_f(2,1,0)-\mu\cdot\tilde{\mu}_f(1,1,0), && \sigma_{33}=\tilde{\mu}_f(2,2,0)-\tilde{\mu}_f^2(1,1,0),\\
		&\sigma_{14}=\tilde{\mu}_f(2,0,1)-\mu\cdot\tilde{\mu}_f(1,0,1), && \sigma_{34}=\tilde{\mu}_f(2,1,1)-\tilde{\mu}_f(1,1,0)\cdot\tilde{\mu}_f(1,0,1),\\
		&\sigma_{22}=\tilde{\mu}_f(0,0,2)-\tilde{\mu}_f^2(0,0,1), && \sigma_{44}=\tilde{\mu}_f(2,0,2)-\tilde{\mu}_f^2(1,0,1).
	\end{align*}
\end{satz}
The proof of Theorem~\ref{CLT_Nb} is provided by Appendix~\ref{Proof of Theorem CLT_Nb}.

\smallskip
In the second step of deriving the Stein-MM estimators' asymptotics, we define the function
\begin{itemize}
	\item $g_\nu(x_1,x_2,x_3,x_4)\coloneqq x_1(x_4-x_3)/(x_3-x_1x_2)$ for~$	\hat{\nu}_{f,\nb}$, 
	\item $g_\pi(x_1,x_2,x_3,x_4)\coloneqq (x_4-x_3)/(x_4-x_1x_2)$ for~$\hat{\pi}_{f,\nb}$.
\end{itemize}
Then, $ \hat{\nu}_{f,\nb}=g_\nu(\overline{\Z}) $, $ \nu=g_\nu(\bmu_Z) $, and $ \hat{\pi}_{f,\nb}=g_\pi(\overline{\Z}) $, $ \pi=g_\pi(\bmu_Z) $, respectively, holds. Applying the Delta method \citep{serfling80} to Theorem~\ref{CLT_Nb}, the following theorems follow.

\begin{satz}\label{Ansu}
	Let $X_1,\ldots,X_n$ be \iid\ according to $ \nb(\nu,\, \tfrac{\nu}{\nu+\mu})  $, and define $ \eta_1\coloneqq  \tilde{\mu}_f(1,1,0)-\mu\cdot\tilde{\mu}_f(0,0,1)$.
	 Then,  $ \hat{\nu}_{f,\nb}$ is asymptotically normally distributed, where the asymptotic variance and bias, respectively, are given by 
	{\small%
	\begin{align*}
		\sigma&_{f,\nb}^2=
		\frac{1}{\eta_1^4n} \Bigg[\mu ^4 \Bigg(\tilde{\mu}_f(0,0,1) \bigg(\tilde{\mu}_f(0,0,1) (\tilde{\mu}_f(2,0,2)-2 \tilde{\mu}_f(2,1,1)+\tilde{\mu}_f(2,2,0))
		\\[1ex]&\quad-2 (\tilde{\mu}_f(1,0,1)-\tilde{\mu}_f(1,1,0)) (\tilde{\mu}_f(1,0,2)-\tilde{\mu}_f(1,1,1)) \bigg)\\
		&+\tilde{\mu}_f(0,0,2) \Big(\tilde{\mu}_f(1,0,1)-\tilde{\mu}_f(1,1,0)\Big)^2\Bigg)\\
		&+2 \mu ^3 \Bigg(\tilde{\mu}_f(0,0,1) \tilde{\mu}_f(1,0,1) \big(\tilde{\mu}_f(2,1,1)-\tilde{\mu}_f(2,2,0)\big)\\
		&-\tilde{\mu}_f(1,1,0) \Big(\tilde{\mu}_f(0,0,1) \big(\tilde{\mu}_f(2,0,2)-\tilde{\mu}_f(2,1,1)\big)+\tilde{\mu}_f(1,0,2) \tilde{\mu}_f(1,1,0)\Big)\\
		&-\tilde{\mu}_f^2(1,0,1) \tilde{\mu}_f(1,1,1)+\tilde{\mu}_f(1,0,1) \tilde{\mu}_f(1,1,0) \big(\tilde{\mu}_f(1,0,2)+\tilde{\mu}_f(1,1,1)\big)\Bigg)\\
		&+\mu ^2 \Bigg(\tilde{\mu}_f(1,1,0) \bigg(2 \tilde{\mu}_f(1,0,1) \Big(-\tilde{\mu}_f(0,0,1) \tilde{\mu}_f(2,0,1)+\tilde{\mu}_f(0,0,1) \tilde{\mu}_f(2,1,0)\\
		&+\tilde{\mu}_f^2(1,1,0)-\tilde{\mu}_f(2,1,1)\Big)+\tilde{\mu}_f(1,1,0) (2 \tilde{\mu}_f(0,0,1) \tilde{\mu}_f(2,0,1)-2 \tilde{\mu}_f(0,0,1) \tilde{\mu}_f(2,1,0)\\
		&+\tilde{\mu}_f(2,0,2))+2 \tilde{\mu}_f^3(1,0,1)-4 \tilde{\mu}_f^2(1,0,1) \tilde{\mu}_f(1,1,0)\bigg)+\tilde{\mu}_f^2(1,0,1) \tilde{\mu}_f(2,2,0)\Bigg)\\
		&-2 \mu\, \tilde{\mu}_f(1,1,0) \big(\tilde{\mu}_f(1,0,1)-\tilde{\mu}_f(1,1,0)\big) \Big(\tilde{\mu}_f(1,0,1) \tilde{\mu}_f(2,1,0)-\tilde{\mu}_f(1,1,0) \tilde{\mu}_f(2,0,1)\Big)\\
		&+\tilde{\mu}_f^2(1,1,0) \sigma ^2 \big(\tilde{\mu}_f(1,0,1)-\tilde{\mu}_f(1,1,0)\big)^2 \Bigg]
	\end{align*}}
and
{\small%
	\begin{align*}
		\mathbb{B}&_{f,\nb}=
		-\frac{1}{\eta_1^3\, n}\Bigg[ \mu ^3 \bigg(\tilde{\mu}_f(0,0,1) \big(\tilde{\mu}_f(1,0,2)-\tilde{\mu}_f(1,1,1)\big)+\tilde{\mu}_f(0,0,2) \big(\tilde{\mu}_f(1,1,0)\\
		&-\tilde{\mu}_f(1,0,1)\big)\bigg)-\mu ^2 \bigg(\tilde{\mu}_f(0,0,1) \big(\tilde{\mu}_f(2,1,1)-\tilde{\mu}_f(2,2,0)\big)-2 \tilde{\mu}_f(1,0,1) \tilde{\mu}_f(1,1,1)\\
		&+\tilde{\mu}_f(1,1,0) \big(\tilde{\mu}_f(1,0,2)+\tilde{\mu}_f(1,1,1)\big)\bigg)\\
		&+\mu  \bigg(\tilde{\mu}_f(1,0,1) \Big(\tilde{\mu}_f(0,0,1) \tilde{\mu}_f(2,1,0)+2 \tilde{\mu}_f^2(1,1,0)-\tilde{\mu}_f(2,2,0)\Big)\\
		&+\tilde{\mu}_f(1,1,0) \Big(\tilde{\mu}_f(0,0,1) \tilde{\mu}_f(2,0,1)-2 \tilde{\mu}_f(0,0,1) \tilde{\mu}_f(2,1,0)+\tilde{\mu}_f(2,1,1)\Big)\\
		&-2 \tilde{\mu}_f^2(1,0,1) \tilde{\mu}_f(1,1,0)\bigg)+\tilde{\mu}_f(1,1,0) \bigg(\tilde{\mu}_f(0,0,1) \sigma ^2 (\tilde{\mu}_f(1,1,0)-\tilde{\mu}_f(1,0,1))\\
		&+\tilde{\mu}_f(1,0,1) \tilde{\mu}_f(2,1,0)-\tilde{\mu}_f(1,1,0) \tilde{\mu}_f(2,0,1)\bigg)\Bigg].
	\end{align*}}
\end{satz}
The proof of Theorem~\ref{Ansu} is provided by Appendix~\ref{Proof of Theorem Ansu}. 

\begin{satz}\label{Fati}
	Let $X_1,\ldots,X_n$ be \iid\ according to $ \nb(\tfrac{\pi\,\mu}{1-\pi},\, \pi) $, and define 
	$ \eta_2\coloneqq  \tilde{\mu}_f(1,0,1)-\mu\cdot\tilde{\mu}_f(0,0,1)$.
	 Then,  $ \hat{\pi}_{f,\nb}$ is asymptotically normally distributed, where the asymptotic variance and bias, respectively, are given by 
	{\small%
	\begin{align*}
		\sigma&_{f,\nb}^2=\frac{1}{\eta_2^4\, n} \Bigg[\mu ^2 \bigg(\tilde{\mu}_f(0,0,1) \Big(\tilde{\mu}_f(0,0,1) \big(\tilde{\mu}_f(2,0,2)-2 \tilde{\mu}_f(2,1,1)+\tilde{\mu}_f(2,2,0)\big)\\
		&-2 (\tilde{\mu}_f(1,0,1)-\tilde{\mu}_f(1,1,0)) (\tilde{\mu}_f(1,0,2)-\tilde{\mu}_f(1,1,1))\Big)\\
		&+\tilde{\mu}_f(0,0,2) (\tilde{\mu}_f(1,0,1)-\tilde{\mu}_f^2(1,1,0))\bigg)\\
		&-2 \mu  \bigg(\tilde{\mu}_f^2(0,0,1) \big(\tilde{\mu}_f(1,0,1)-\tilde{\mu}_f(1,1,0)\big) \big(\tilde{\mu}_f(2,0,1)-\tilde{\mu}_f(2,1,0)\big)\\
		&-\tilde{\mu}_f(0,0,1) \Big(\tilde{\mu}_f^3(1,0,1)-2 \tilde{\mu}_f^2(1,0,1) \tilde{\mu}_f(1,1,0)+\tilde{\mu}_f(1,0,1) \Big(\tilde{\mu}_f^2(1,1,0)\\
		&+\tilde{\mu}_f(2,1,1)-\tilde{\mu}_f(2,2,0)\Big)+\tilde{\mu}_f(1,1,0) \big(\tilde{\mu}_f(2,1,1)-\tilde{\mu}_f(2,0,2)\big)\Big)\\
		&+\big(\tilde{\mu}_f(1,0,1)-\tilde{\mu}_f(1,1,0)\big) \big(\tilde{\mu}_f(1,0,1) \tilde{\mu}_f(1,1,1)-\tilde{\mu}_f(1,0,2) \tilde{\mu}_f(1,1,0)\big)\bigg)\\
		&+\tilde{\mu}_f^2(0,0,1) \sigma ^2 \big(\tilde{\mu}_f(1,0,1)-\tilde{\mu}_f(1,1,0)\big)^2+\tilde{\mu}_f^2(1,0,1) \tilde{\mu}_f(2,2,0)\\
		&-2 \tilde{\mu}_f(0,0,1) \big(\tilde{\mu}_f(1,0,1)-\tilde{\mu}_f(1,1,0)\big) \big(\tilde{\mu}_f(1,0,1) \tilde{\mu}_f(2,1,0)-\tilde{\mu}_f(1,1,0) \tilde{\mu}_f(2,0,1)\big)\\
		&-2 \tilde{\mu}_f(1,0,1) \tilde{\mu}_f(1,1,0) \tilde{\mu}_f(2,1,1)+\tilde{\mu}_f^2(1,1,0) \tilde{\mu}_f(2,0,2)\Bigg]
	\end{align*}}
	and
	{\small%
	\begin{align*}
		\mathbb{B}&_{f,\nb}=-\frac{1}{\eta_2^3\, n} \Bigg[\mu^2 \bigg(\tilde{\mu}_f(0,0,1) \big(\tilde{\mu}_f(1,0,2)-\tilde{\mu}_f(1,1,1)\big)\\
		&+\tilde{\mu}_f(0,0,2) \big(\tilde{\mu}_f(1,1,0)-\tilde{\mu}_f(1,0,1)\big)\bigg)\\
		&+\mu \bigg(\tilde{\mu}_f^2(0,0,1) \big(\tilde{\mu}_f(2,0,1)-\tilde{\mu}_f(2,1,0)\big)\\
		&+\tilde{\mu}_f(0,0,1) \Big(\tilde{\mu}_f(1,0,1) \big(\tilde{\mu}_f(1,1,0)-\tilde{\mu}_f(1,0,1)\big)-\tilde{\mu}_f(2,0,2)+\tilde{\mu}_f(2,1,1)\Big)\\
		&+\tilde{\mu}_f(1,0,1) \big(\tilde{\mu}_f(1,0,2)+\tilde{\mu}_f(1,1,1)\big)-2 \tilde{\mu}_f(1,0,2) \tilde{\mu}_f(1,1,0)\bigg)\\
		&+\tilde{\mu}_f(1,1,0) \Big(\tilde{\mu}_f^2(0,0,1)\, \sigma^2-2 \tilde{\mu}_f(0,0,1) \tilde{\mu}_f(2,0,1)+\tilde{\mu}_f(2,0,2)\Big)\\
		&+\tilde{\mu}_f(0,0,1) \tilde{\mu}_f(1,0,1) \Big(-\tilde{\mu}_f(0,0,1)\,  \sigma^2+\tilde{\mu}_f(2,0,1)+\tilde{\mu}_f(2,1,0)\Big)\\
		&-\tilde{\mu}_f^3(1,0,1)+\tilde{\mu}_f^2(1,0,1) \tilde{\mu}_f(1,1,0)-\tilde{\mu}_f(1,0,1) \tilde{\mu}_f(2,1,1) \Bigg].
	\end{align*}}
\end{satz}
The proof of Theorem~\ref{Fati} is provided by Appendix~\ref{Proof of Theorem Fati}. 

\bigskip
Our first special case shall be the function $f_\alpha(x)=\alpha^x$ with $\alpha\in (0;1)$, which is inspired by \citet{Kemp87}. For evaluating the asymptotics in Theorems~\ref{CLT_Nb}--\ref{Fati}, we need to compute the moments
$$
\tilde{\mu}_{f_\alpha}(k,l,m)
= \e[X^k\cdot f_\alpha(X)^l\cdot f_\alpha(X+1)^m]
= \alpha^m\, \e[X^k\cdot \alpha^{(l+m) X}].
$$
As shown in the following, this can be done by explicit closed-form expressions. The idea is to utilize the probability generating function (pgf) of the NB-distribution,
$$
\pgf(z) \coloneqq \e[z^X]\ =\ \bigg(\frac{\pi}{1-(1-\pi)z} \bigg)^\nu,
$$
together with the following property:
$$
\e\big[X_{(k)}\, z^X\big]\ =\ z^k\cdot\frac{d^k}{dz^k}\,\pgf(z),
$$
where $ x_{(r)}\coloneqq x\cdots(x-r+1) $ for $r\in\mathbb{N}_0$ denote the falling factorials. The main result is summarized by the following lemma.

\begin{lemma}\label{facZx}
	Let $X\sim\nb(\nu,\,\pi)$. For the mixed factorial moments, we have
$$
		 \e[X_{(k)}z^X]
		\ =\ \frac{(1-\pi)^k\, (\nu+k-1)_{(k)}\, z^k}{\big(1-(1-\pi)z\big)^k}\, \pgf(z).
$$
\end{lemma}
The proof of Lemma~\ref{facZx} is provided by Appendix~\ref{Proof of Lemma facZx}. The factorial moments are easily transformed into raw moments by using the relation $x^k=\sum_{j=0}^{k} S_{k,j}\, x_{(j)}$, where $S_{k,j}$ are the Sterling numbers of the second kind \citep[see][p.~12]{Johnson05}. Then, $\e[X^k\cdot \alpha^{(l+m) X}]$ follows by plugging-in $z=\alpha^{l+m}$ into Lemma~\ref{facZx}.

\begin{figure}[th!]
	\center\footnotesize
	(a)\hspace{-3ex}\includegraphics[scale=0.5, viewport=0 45 335 305, clip=]{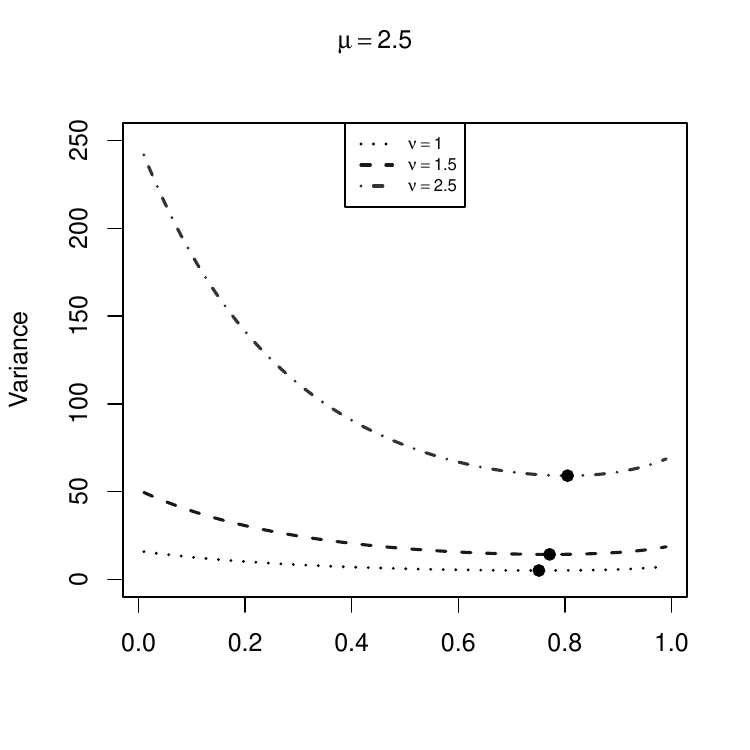}~$\alpha$
	\qquad
	(b)\hspace{-3ex}\includegraphics[scale=0.5, viewport=0 45 335 305, clip=]{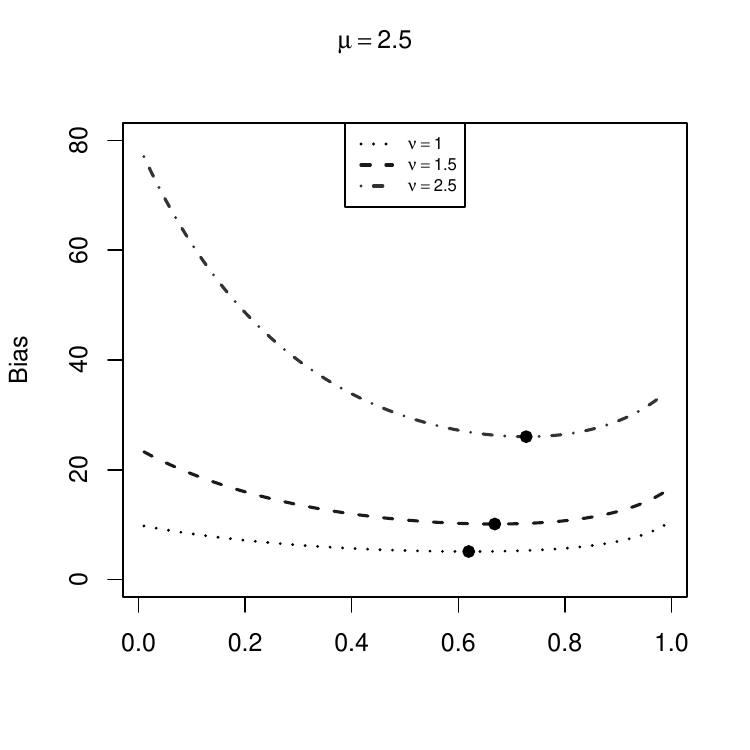}~$\alpha$
	\\[2ex]
	(c)\hspace{-3ex}\includegraphics[scale=0.5, viewport=0 45 335 305, clip=]{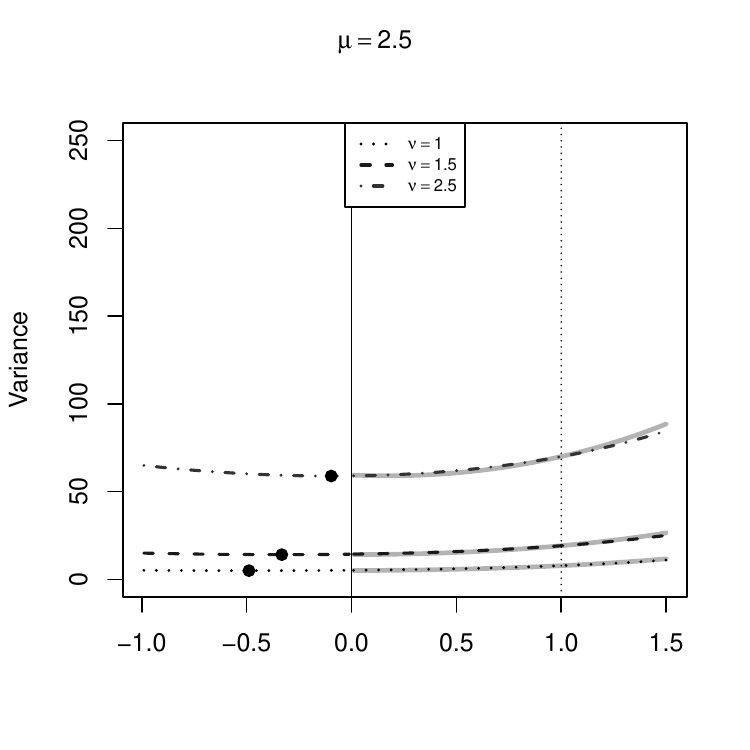}~$a$
	\qquad
	(d)\hspace{-3ex}\includegraphics[scale=0.5, viewport=0 45 335 305, clip=]{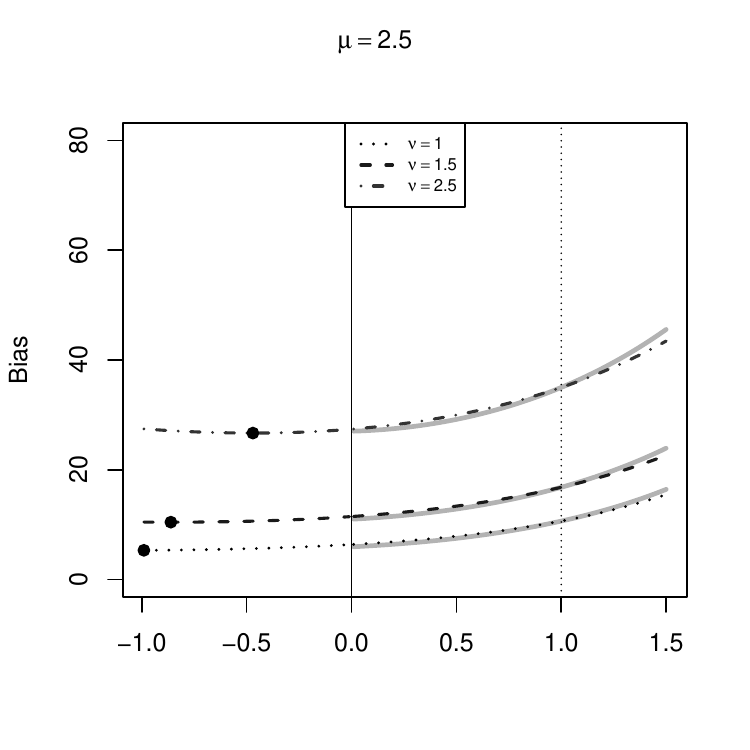}~$a$
	
	\caption{Stein-MM estimator $ \hat{\nu}_{f,\nb}$ for $\mu=2.5$. Plots of $n\,\sigma^2 _{f,\nb} $ and $ n\,\mathbb{B}_{f,\nb} $ for parametrization \eqref{NB1}, where points indicate minimal variance and bias values. Weighting function (a)--(b) $f(x)=\alpha^x$ with $\alpha\in(0,1)$, and (c)--(d) $f(x)=(x+1)^a$ with $a\in(-1,1.5)$. The gray graphs in (c)--(d) correspond to the comparative choice $f(x)=x^a$, which leads to the default MM estimator for $a=1$ (dotted lines).}
	\label{fig_VarBias_Nb_nu}
\end{figure}

\begin{figure}[th!]
	\center\footnotesize
	(a)\hspace{-3ex}\includegraphics[scale=0.5, viewport=0 45 335 305, clip=]{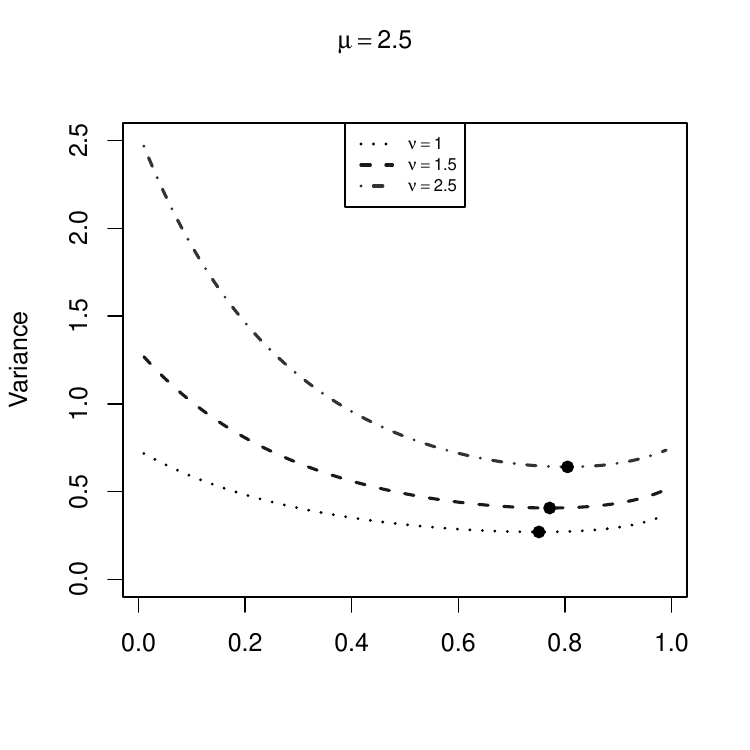}~$\alpha$
	\qquad
	(b)\hspace{-3ex}\includegraphics[scale=0.5, viewport=0 45 335 305, clip=]{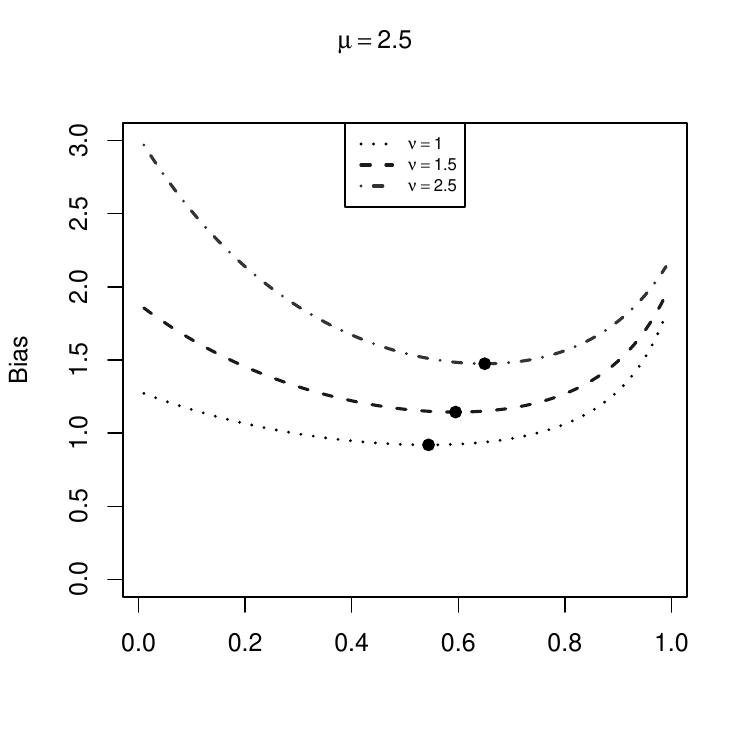}~$\alpha$
	\\[2ex]
	(c)\hspace{-3ex}\includegraphics[scale=0.5, viewport=0 45 335 305, clip=]{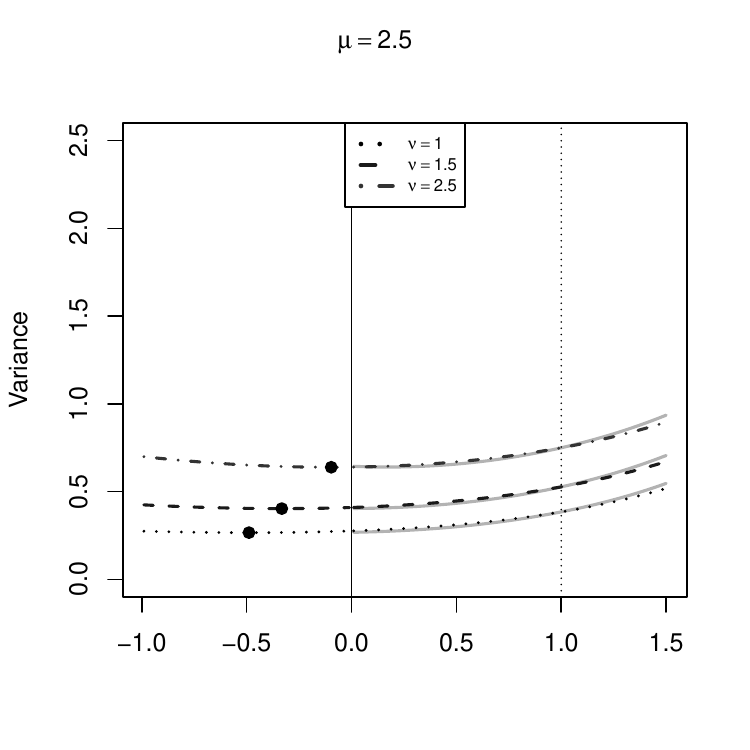}~$a$
	\qquad
	(d)\hspace{-3ex}\includegraphics[scale=0.5, viewport=0 45 335 305, clip=]{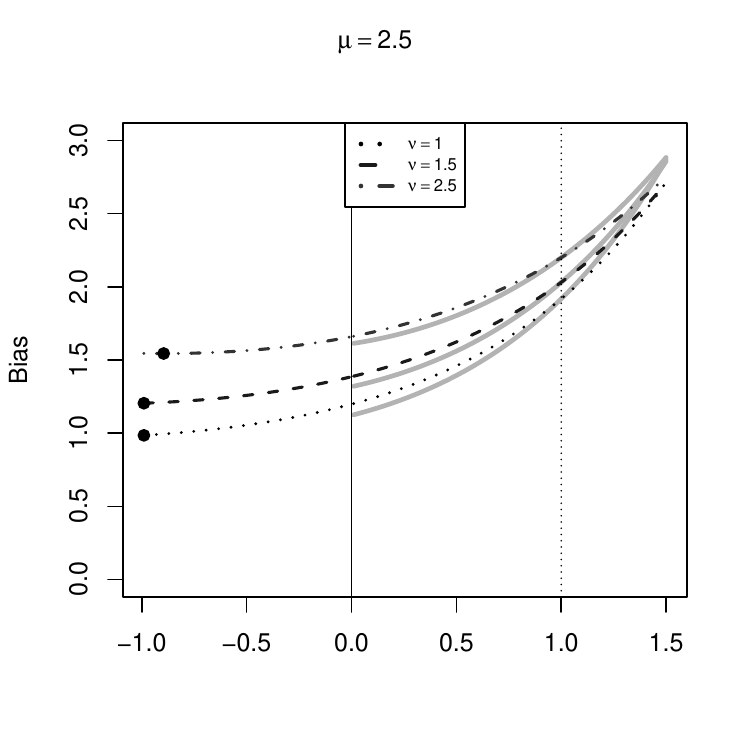}~$a$
	
	\caption{Stein-MM estimator $ \hat{\pi}_{f,\nb}$ for $\mu=2.5$. Plots of $n\,\sigma^2 _{f,\nb} $ and $ n\,\mathbb{B}_{f,\nb} $ for parametrization \eqref{NB2}, where points indicate minimal variance and bias values. Weighting function (a)--(b) $f_\alpha(x)=\alpha^x$ with $\alpha\in(0,1)$, and (c)--(d) $f_{a,1}(x)=(x+1)^a$ with $a\in(-1,1.5)$. The gray graphs in (c)--(d) correspond to the comparative choice $f_a(x)=x^a$, which leads to the default MM estimator for $a=1$ (dotted lines).}
	\label{fig_VarBias_Nb_pi}
\end{figure}

\smallskip
While general closed-form formulae are possible in this way for $\tilde{\mu}_{f_\alpha}(k,l,m)$ as well as for Theorems~\ref{CLT_Nb}--\ref{Fati}, the obtained results are very complex such that we decided to omit the final expressions. Instead, we compute $\tilde{\mu}_{f_\alpha}(k,l,m)$ and, thus, the expressions of Theorems~\ref{CLT_Nb}--\ref{Fati} numerically. This is easily done in practice, in fact for any reasonable choice of the function~$f$, by computing
$$
\tilde{\mu}_{f_\alpha}(k,l,m)
\ \approx\ \sum_{x=0}^M x^k\cdot f(x)^l\cdot f(x+1)^m\cdot P(X=x),
$$
where the upper summation limit~$M$ is chosen sufficiently large, \eg such that $M^k\cdot f(M)^l\cdot f(M+1)^m\cdot P(X=M)$ falls below a specified tolerance limit. In this way, we generated the illustrative graphs in Figures~\ref{fig_VarBias_Nb_nu} (estimator $\hat{\nu}_{f,\nb}$) and~\ref{fig_VarBias_Nb_pi} (estimator $\hat{\pi}_{f,\nb}$). There, parts~(a)--(b) always refer to the above choice $f_\alpha(x)=\alpha^x$, and clear minima for variance and bias for $f_\alpha(x)=\alpha^x$ can be recognized. To be able to compare with the respective default MM estimator, we did analogous computations for $f_a(x)=x^a$ (where $a=1$ for the default MM estimator), which, however, is only defined for $a>0$ as~$X$ becomes zero with positive probability. As can be seen from the gray curves in parts~(c)--(d), variance and basis usually do not attain a local minimum for $a>0$. Therefore, parts~(c)--(d) mainly focus on a slight modification of the weight function, namely $f_{a,1}(x)=(x+1)^a$, which is also well-defined for $a<0$. 

\begin{table}[!t]
	\centering
	\caption{Stein-MM estimators $ \hat{\nu}_{f,\nb}$ and $ \hat{\pi}_{f,\nb}$, optimal choices for $\alpha$ or~$a$, respectively (columns ``$a_{\textup{opt}}$''), and corresponding minimal value of variance and bias (columns ``min.'')}
	\label{tabNBoptimal}
	
	\smallskip
	\resizebox{\linewidth}{!}{
\begin{tabular}{lcc|rr@{\qquad}rr|rr@{\qquad}rr}
\toprule
 &  &  & \multicolumn{4}{l|}{$ \hat{\nu}_{f,\nb}$, optimal choice for} & \multicolumn{4}{l}{$ \hat{\pi}_{f,\nb}$, optimal choice for} \\
\multicolumn{3}{l|}{$\mu=2.5$} & \multicolumn{2}{l}{variance} & \multicolumn{2}{l|}{bias} & \multicolumn{2}{l}{variance} & \multicolumn{2}{l}{bias} \\
$f(x)$ & $\nu$ & $\pi$ & $a_{\textup{opt}}$ & min. & $a_{\textup{opt}}$ & min. & $a_{\textup{opt}}$ & min. & $a_{\textup{opt}}$ & min. \\
\midrule
$\alpha^x$ & 1.0 & 0.286 & 0.751 & 5.095 & 0.620 & 5.093 & 0.751 & 0.271 & 0.544 & 0.920 \\
 & 1.5 & 0.375 & 0.771 & 14.271 & 0.668 & 10.102 & 0.771 & 0.407 & 0.595 & 1.144 \\
 & 2.5 & 0.500 & 0.805 & 59.113 & 0.727 & 26.022 & 0.805 & 0.641 & 0.650 & 1.475 \\
\midrule
$(x+1)^a$ & 1.0 & 0.286 & -0.489 & 5.002 & -0.990 & 5.311 & -0.489 & 0.267 & -0.990 & 0.985 \\
 & 1.5 & 0.375 & -0.332 & 14.130 & -0.861 & 10.440 & -0.332 & 0.404 & -0.990 & 1.205 \\
 & 2.5 & 0.500 & -0.097 & 58.908 & -0.470 & 26.680 & -0.097 & 0.639 & -0.896 & 1.544 \\
\bottomrule
\end{tabular}}
\end{table}

\smallskip
The optimal choices for~$\alpha$ and~$a$, respectively, lead to very similar variance and bias values, see Table~\ref{tabNBoptimal}. While $\alpha^x$ leads to a slightly larger variance than $(x+1)^a$, its optimal bias is visibly lower. 
For both choices of~$f$, however, the optimal Stein-MM estimators perform clearly better than the default MM estimator, see the dotted line at $a=1$ in parts (c)--(d) in Figures~\ref{fig_VarBias_Nb_nu} and~\ref{fig_VarBias_Nb_pi}. Altogether, also in view of the fact that explicit closed-form expressions are possible for $\alpha^x$ (although being rather complex), we prefer to use $f_\alpha(x)=\alpha^x$ as the weighting function, in accordance to \citet{Kemp87}. 
For this choice, we also did a simulation experiment with $10^5$ replications (in analogy to Remark~\ref{bemIGsimulation}), in order to check the finite-sample performance of the asymptotic expressions for variance and bias. We generally observed a very good agreement between asymptotic and simulated values. Especially for the estimator~$\hat{\pi}_{f_\alpha,\nb}$, the asymptotic approximations show an excellent performance, whereas the estimator~$\hat{\nu}_{f_\alpha,\nb}$ sometimes leads to extreme estimates if $\nu=2.5$ and $n=100$. But except these few outlying estimates, also~$\hat{\nu}_{f_\alpha,\nb}$ is well described by the asymptotic formulae. Detailed simulation results are available from the authors upon request.

	\begin{table}[t]
		\centering
		\caption{Counts of red mites on apple leaves from Example~\ref{bspNbdata1}: Stein-MM estimates $\hat{\nu}_{f_\alpha,\nb}$ (upper part) and  $\hat{\pi}_{f_\alpha,\nb}$ (lower part) for different choices of function $f_\alpha(x)=\alpha^x$.}
		\label{Mites_data1}
		
		\smallskip
			$\displaystyle
			\begin{array}{r|c@{\qquad}c@{\qquad}c@{\qquad}c@{\qquad}c|c}
				\toprule
				\alpha & 0.25 & 0.5 & 0.530 & 0.690 & 0.75 & \text{---} \\ 
				\text{\footnotesize Notes} &  & & \text{\footnotesize (i)} & \text{\footnotesize (ii)} &  & \text{\footnotesize (iii)}     \\
				\midrule
				\hat{\nu}_{f_\alpha,\nb} & 0.967 &0.963 &0.967&1.009& 1.032& 1.167 \\
				\bottomrule
				\multicolumn{2}{l}{\text{\footnotesize (i)\quad $\mathbb{B}_{f_\alpha,\nb}$-optimal}} &
				\multicolumn{2}{l}{\text{\footnotesize (ii)\quad $\sigma^2 _{f_\alpha,\nb}$-optimal}} &
				\multicolumn{2}{l}{\text{\footnotesize (iii)\quad MM estimate}} 
			\end{array}
			$
		
		\vspace{3ex}
			$\displaystyle
			\begin{array}{r|c@{\qquad}c@{\qquad}c@{\qquad}c@{\qquad}c|c}
				\toprule
				\alpha & 0.222&0.25&0.5&0.690& 0.75& \text{---} \\ 
				\text{\footnotesize Notes} &   \text{\footnotesize (i)}& & & \text{\footnotesize (ii)} &  & \text{\footnotesize (iii)}     \\
				\midrule
				\hat{\pi}_{f_\alpha,\nb} &  0.459 &0.457& 0.456&0.468& 0.474& 0.504\\
				\bottomrule
				\multicolumn{2}{l}{\text{\footnotesize (i)\quad $\mathbb{B}_{f_\alpha,\nb}$-optimal}} &
				\multicolumn{2}{l}{\text{\footnotesize (ii)\quad $\sigma^2 _{f_\alpha,\nb}$-optimal}} &
				\multicolumn{2}{l}{\text{\footnotesize (iii)\quad MM estimate}} 
			\end{array}
			$
	\end{table}

\begin{bsp}\label{bspNbdata1}
As an illustrative data example, let us consider $n=150$ counts of red mites on apple leaves \citep[see][p.~271]{RuedOr99}, who confirmed ``a good fit of the negative binomial'' for these data. The parameter~$\mu$ is estimated by the sample mean as~$\approx 1.147$. 
In case of the $(\mu, \nu)$-parametrization, we use the ordinary MM~estimator as an initial estimator for~$\nu$, leading to the value $\approx 1.167$. Based on this initial model fit, we search for Stein-MM estimators with $f_\alpha(x)=\alpha^x$ having an improved performance. The resulting estimates (together with further estimates for comparative purposes) are summarized in the upper part of Table~\ref{Mites_data1}. It can be seen that the initial estimate (last column) is corrected downwards to a value close to~1 (\ie we essentially end up with the special case of a geometric distribution). Here, it is interesting to note that the numerically computed ML estimate as reported in \citet{RuedOr99}, also leads to such a result, namely to the value~1.025. In this context, we also recall \citet{Kemp87} who proposed the choice $f_\alpha(x)=\alpha^x$ to get a closed-form approximate ML~estimator for~$\nu$.

\smallskip
We repeated the aforementioned estimation procedure also for the $(\mu, \pi)$-paramete\-rization, starting with the initial MM-estimate $\approx 0.504$ for~$\pi$, see the lower part of Table~\ref{Mites_data1}. Again, the initial estimate is corrected downwards to a value around~0.46. 
\end{bsp}

\section{Conclusions}
\label{Conclusions}
%
In this article, we demonstrated how Stein characterizations of (continuous or discrete) distributions can be utilized to derive improved moment estimators of model parameters. The main idea is to choose an appropriate type of weighting function such that the resulting Stein-MM estimator has lower variance and bias than existing estimators. Here, the choice of the weighting function can be done based on asymptotic distributions: as the Stein-MM estimators are given by closed-form expressions with asymptotic normality, one can easily derive an optimal choice from a given family of weighting functions. This procedure was exemplified for three types of distribution: the exponential distribution, the inverse Gaussian distribution, and the negative-binomial distribution. For all these distribution families, we observed an appealing performance in various aspects, and we also demonstrated the application of our findings to real-world data examples. 

\smallskip
Our research also gives rise to several directions for future research. While our main focus was on selecting the weight function with respect to minimal bias or variance, we also briefly pointed out in Section~\ref{Stein Estimation of Exponential Distribution} that such a choice could also be motivated by robustness to outliers. In fact, there are some analogies to ``M-estimation'' as introduced by \citet{huber64}. It appears to be promising to analyze if robustified MM~estimators can be achieved by suitable classes of weighting function. As another direction for future research (briefly sketched in Section~\ref{Stein Estimation of Exponential Distribution}), the performance of GoF-tests based on Stein-MM estimators should be investigated. Finally, one should analyze Stein-MM estimators in a regression or time-series context.

\subsubsection*{Acknowledgments}
This research was funded by the Deutsche Forschungsgemeinschaft (DFG, German Research Foundation) -- Projektnummer 437270842.



\appendix
\small
\numberwithin{equation}{section}
\numberwithin{table}{section}

%
\section{Derivations}\label{Derivations}
%

%
\subsection{Proof of Theorem~\ref{CLT_Exp}}\label{Proof of Theorem CLT_Exp}
%
The asymptotic normality immediately follows from the Lindeberg--L\'evy CLT \citep[p.~28]{serfling80}. 
	For the covariances, we get 
	\begin{align*}
		\sigma_{11}&=CoV[f'(X_i),f'(X_i)]=\e[f'(X_i)^2]-\e[f'(X_i)]^2\\
		&\overset{\eqref{Lionel}}{=}\e[f'(X_i)^2]-\lambda^2\,\e[f(X_i)]^2=\mu_f(0,0,2)-\lambda^2\cdot\mu_f^2(0,1,0);\\
		\sigma_{22}&=CoV[f(X_i),f(X_i)]=\e[f(X_i)^2]-\e[f(X_i)]^2=\mu_f(0,2,0)-\mu_f^2(0,1,0);\\
		\sigma_{12}&=CoV[f'(X_i),f(X_i)]=\e[f'(X_i)f(X_i)]-\e[f'(X_i)]\,\e[f(X_i)]\\
		&\overset{\eqref{Lionel}}{=}\tfrac{\lambda}{2}\,\e[f(X_i)^2]-\lambda\,\e[f(X_i)]^2=\tfrac{\lambda}{2}\big(\sigma_{22}-\mu_f^2(0,1,0)\big).
	\end{align*}
	In the last line, we applied the Stein-identity \eqref{Lionel} to $ g(X_i)\coloneqq\tfrac{1}{2}f(X_i)^2 $ with the derivative $ g'(X_i)=f'(X_i)f(X_i) $: 
	\begin{align*}
		\e[f'(X_i)f(X_i)]=\e[g'(X_i)]=\lambda\e[g(X_i)]=\tfrac{\lambda}{2}\e[f(X_i)^2].
	\end{align*}
	We conclude the proof by noting that the second expression for~$\sigma_{12}$ in Theorem~\ref{CLT_Exp} immediately follows by using the expression for~$\sigma_{22}$.
%

\subsection{Proof of Theorem~\ref{Diego}}\label{Proof of Theorem Diego}
%
First, we evaluate the gradient and Hessian of~$g$ in~$\bmu_Z$, which leads to $ \D $ and $ \Hes $ given by
\begin{align*}
	\D=\frac{1}{\mu_f(0,1,0)}\big(1,-\lambda\big)^\top, \qquad \Hes=\frac{1}{\mu_f^2(0,1,0)}\begin{pmatrix} 
		0 & -1 \\ 
		-1 & 2\lambda \\
	\end{pmatrix}.
\end{align*}
Applying the Delta method to Theorem~\ref{CLT_Exp}, the asymptotic normality 
\begin{align*}
	\sqrt{n}\big(\hat{\lambda}_{f,\Exp}-\lambda\big) \ \xrightarrow{\text{d}}\ \norm\big(0, \sigma^2\big)\  \text{with}\  \sigma^2=\D\bSigma\D^\top = d_1^2\sigma_{11}+d_2^2\sigma_{22}+2d_1d_2\sigma_{12}
\end{align*}
follows. The 2nd-order Taylor approximation $ \hat{\lambda}_{f,\Exp} - \lambda\approx \D(\overline{\Z}-\bmu_Z)+\tfrac{1}{2}(\overline{\Z}-\bmu_Z)^\top\Hes(\overline{\Z}-\bmu_Z)$ allows to conclude the asymptotic bias as
\begin{align*}
	\mathbb{B}_{f,\Exp}&=\tfrac{1}{n} \Big( \tfrac{1}{2}h_{22}\sigma_{22} +h_{12}\sigma_{12}\Big).
\end{align*}
The explicit expression for the asymptotic variance $\sigma_{f,\Exp}^2 = \frac{\sigma^2}{n}$ follows by applying $ \D $ and Theorem \ref{CLT_Exp}:
	\begin{align*}
		\sigma_{f,\Exp}^2
		&=\frac{1}{n}\,\frac{1}{\mu_f^2(0,1,0)}\,\Bigg[\sigma_{11}+\lambda^2\,\sigma_{22} -2\lambda\cdot\tfrac{\lambda}{2}\Big(\sigma_{22}-\mu_f^2(0,1,0)\Big)\Bigg]\\
		&=\frac{1}{n}\,\frac{1}{\mu_f^2(0,1,0)}\,\Bigg[\sigma_{11} +\lambda^2\mu_f^2(0,1,0)\Bigg]
		=\frac{1}{n}\cdot\frac{\mu_f(0,0,2)}{\mu_f^2(0,1,0)}.
	\end{align*}
Similarly, using~$\Hes$, the asymptotic bias becomes
	\begin{align*}
		\mathbb{B}_{f,\Exp}&=\frac{1}{n}\Bigg[ \frac{1}{2}\cdot\frac{2\lambda\cdot\sigma_{22}}{\mu_f^2(0,1,0)} - \frac{\lambda\cdot\sigma_{22}-\tfrac{\lambda}{2}\, \mu_f(0,2,0)}{\mu_f^2(0,1,0)}\Bigg]
		=\frac{\lambda}{2n}\cdot\frac{\mu_f(0,2,0)}{\mu_f^2(0,1,0)}.
	\end{align*}
This completes the proof of Theorem~\ref{Diego}.

%
\subsection{Proof of Corollary~\ref{korExpAsymXa}}\label{Proof of Corollary korExpAsymXa}
%
We start by proving \eqref{ExpMomXa}. As the pdf of~$X$ is given by $\phi(x)=\lambda e^{-\lambda x}$ for $x > 0 $ and zero otherwise, we get
	\begin{align*}
		\e[X^a]&=\int_{0}^{\infty}x^a \phi(x)\diff x=\int_{0}^{\infty}x^a \lambda e^{-\lambda x}\diff x
		\\
		&=\frac{\Gamma(a+1)}{\lambda^a}\underbrace{\int_{0}^{\infty}x^a \frac{\lambda^{a+1}}{\Gamma(a+1)} e^{-\lambda x}\diff x}_{=1}=\frac{\Gamma(a+1)}{\lambda^a},
	\end{align*}
	where the integrand of the last integral is the pdf of a two-parameter gamma distribution, namely $ \gam(a+1,\tfrac{1}{\lambda}) $ with $ a>-1$ and $\lambda>0 $ according to \citet[p.~343]{Johnson95}. Note that $\e[X^a]$ corresponds to $ \mu_{f_a}(0,1,0)$ in our notation. In view of Theorem \ref{Diego}, we also need the following moments, which are implied by \eqref{ExpMomXa}:
\begin{align*}
\mu_{f_a}(0,2,0)&=\e\big[X^{2a}\big]=\frac{\Gamma(2a+1)}{\lambda^{2a}}
\quad\text{if } a>-\tfrac{1}{2}, \\ 
\mu_{f_a}(0,0,2)&=\e\big[(aX^{a-1})^2\big]=a^2\,\e\big[X^{2(a-1)}\big]=a^2\,\frac{\Gamma(2a-1)}{\lambda^{2(a-1)}}
\quad\text{if } a>\tfrac{1}{2}.
	\end{align*}
Inserting these expressions into Theorem \ref{Diego}, we get
	\begin{align*}
		\sigma_{f_a,\Exp}^2&=\frac{1}{n}\cdot\frac{\mu_{f_a}(0,0,2)}{\mu_{f_a}^2(0,1,0)}
		=\frac{a^2}{n}\frac{\lambda^{2a}}{\lambda^{2(a-1)}}\,\frac{\Gamma(2a-1)}{\Gamma(a+1)^2}
		=\frac{\lambda^2}{n}\, \frac{\Gamma(2a-1)}{\Gamma(a)^2}
		=\frac{\lambda^2}{n}\binom{2(a-1)}{a-1},
	\\[0.5cm]
		\mathbb{B}_{f_a,\Exp}&=\frac{\lambda}{2n}\cdot\frac{\mu_{f_a}(0,2,0)}{\mu_{f_a}^2(0,1,0)}
		=\frac{\lambda}{2n}\frac{\Gamma(2a+1)}{\Gamma(a+1)^2}
		=\frac{\lambda}{2n}\binom{2a}{a}.
	\end{align*}
	This completes the proof of Corollary~\ref{korExpAsymXa}.

%
\subsection{Proof of Corollary~\ref{korExpAsymuX}}\label{Proof of Corollary korExpAsymuX}
%
	Let us start by deriving \eqref{ExpMomuX}. We have 
	\begin{align*}
		\e\big[u^X\big]&=\int_{0}^{\infty}u^x \lambda e^{-\lambda x}\diff x=\lambda\int_{0}^{\infty}e^{x\ln(u)}e^{-\lambda x}\diff x\\
		&=\lambda\int_{0}^{\infty}e^{-x(\lambda-\ln(u))}\diff x=\frac{\lambda}{\lambda-\ln(u)}.
	\end{align*}
	This immediately leads to
	\begin{align*}
\mu_{f_u}(0,1,0)&=\e\big[1-u^X\big]=1-\e\big[u^X\big]=\frac{-\ln(u)}{\lambda-\ln(u)},\\
\mu_{f_u}(0,2,0)&=\e\big[(1-u^X)^2\big]= 1-2\e\big[u^X\big]+\e\big[(u^2)^X\big]\\
&= \frac{-\ln(u)-\lambda}{\lambda-\ln(u)}+\frac{\lambda}{\lambda-2\ln(u)}
=\frac{2\ln(u)^2}{(\lambda-\ln(u))(\lambda-2\ln(u))},\\
\mu_{f_u}(0,0,2)&=\e\big[(\ln(u)\,u^X)^2\big]=\ln(u)^2\,\e\big[(u^2)^X\big]=\frac{\lambda\ln(u)^2}{\lambda-2\ln(u)}.
	\end{align*}
Insertion into Theorem \ref{Diego} leads to
	\begin{align*}
		\sigma_{f_u,\Exp}^2&=\frac{1}{n}\cdot\frac{\mu_{f_u}(0,0,2)}{\mu_{f_u}^2(0,1,0)}=\frac{1}{n}\,\frac{\lambda\ln(u)^2}{\lambda-2\ln(u)}\,\frac{(\lambda-\ln(u))^2}{\ln(u)^2}
		=\frac{\lambda}{n}\frac{(\lambda-\ln(u))^2}{\lambda-2\ln(u)},
		\\[0.5cm]
		\mathbb{B}_{f_u,\Exp}&=\frac{\lambda}{2n}\cdot\frac{\mu_{f_u}(0,2,0)}{\mu_{f_u}^2(0,1,0)}=\frac{\lambda}{2n}\,\frac{2\ln(u)^2}{(\lambda-\ln(u))(\lambda-2\ln(u))}\,\frac{(\lambda-\ln(u))^2}{\ln(u)^2}
		=\frac{\lambda}{n}\frac{\lambda-\ln(u)}{\lambda-2\ln(u)}.
	\end{align*}
Finally, the MSE equals 
\begin{align*}
	\text{MSE}_{f_u,\Exp}&=\sigma_{f_u,\Exp}^2 + \mathbb{B}_{f_u,\Exp}^2
	= \frac{\lambda}{n}\frac{(\lambda-\ln(u))^2}{\lambda-2\ln(u)} + \bigg[\frac{\lambda}{n}\frac{\lambda-\ln(u)}{\lambda-2\ln(u)}\bigg]^2\\
	&=\frac{\lambda}{n}\frac{(\lambda-\ln(u))^2}{(\lambda-2\ln(u))^2}\,\bigg[\lambda-2\ln(u) + \frac{\lambda}{n}\bigg]
	=\frac{\lambda}{n}\frac{(\lambda-\ln(u))^2}{(\lambda-2\ln(u))^2}\,\bigg[\lambda(1+\tfrac{1}{n})-2\ln(u) \bigg].
\end{align*}
This completes the proof of Corollary~\ref{korExpAsymuX}.

%
\subsection{Proof of Theorem~\ref{CLT_IG}}\label{Proof of Theorem CLT_IG}
%
The asymptotic normality immediately follows from the Lindeberg--L\'evy CLT \citep[p.~28]{serfling80}. 
For the covariances, we get 
\begin{align*}
	\sigma_{11}&=CoV[X_i,\, X_i]=\V[X_i]=\mu^3/\lambda;\\
	\sigma_{12}&=CoV[X_i,\, f(X_i)]=\e[X_if(X_i)]-\e[X_i]\,\e[f(X_i)]=\mu_f(1,1,0)-\mu\cdot\mu_f(0,1,0);\\
	\sigma_{13}&=CoV[X_i,\, X_if(X_i)]=\e[X_i^2f(X_i)]-\e[X_i]\,\e[X_if(X_i)]\\
	&=\mu_f(2,1,0)-\mu\cdot\mu_f(1,1,0);\\
	\sigma_{14}&=CoV[X_i,\, X_i^2f(X_i)]=\e[X_i^3f(X_i)]-\e[X_i]\,\e[X_i^2f(X_i)]\\
	&=\mu_f(3,1,0)-\mu\cdot\mu_f(2,1,0);\\
	\sigma_{15}&=CoV[X_i,\, X_i^2f'(X_i)]=\e[X_i^3f'(X_i)]-\e[X_i]\,\e[X_i^2f'(X_i)]\\
	&=\mu_f(3,0,1)-\mu\cdot\mu_f(2,0,1);\\
	\sigma_{22}&=CoV[f(X_i),\, f(X_i)]=\e[f(X_i)^2]-\e[f(X_i)]^2=\mu_f(0,2,0)-\mu_f^2(0,1,0);\\
	\sigma_{23}&=CoV[f(X_i),\, X_if(X_i)]=\e[X_if(X_i)^2]-\e[f(X_i)]\,\e[X_if(X_i)]\\
	&=\mu_f(1,2,0)-\mu_f(0,1,0)\cdot\mu_f(1,1,0);\\
	\sigma_{24}&=CoV[f(X_i),\, X_i^2f(X_i)]=\e[X_i^2f(X_i)^2]-\e[f(X_i)]\,\e[X_i^2f(X_i)]\\
	&=\mu_f(2,2,0)-\mu_f(0,1,0)\cdot\mu_f(2,1,0);\\
	\sigma_{25}&=CoV[f(X_i),\, X_i^2f'(X_i)]=\e[X_i^2f(X_i)f'(X_i)]-\e[f(X_i)]\,\e[X_i^2f'(X_i)]\\&=\mu_f(2,1,1)-\mu_f(0,1,0)\cdot\mu_f(2,0,1);\\
	\sigma_{33}&=CoV[X_if(X_i),\, X_if(X_i)]=\e[X_i^2f(X_i)^2]-\e[X_if(X_i)]^2=\mu_f(2,2,0)-\mu_f^2(1,1,0);\\
	\sigma_{34}&=CoV[X_if(X_i),\, X_i^2f(X_i)]=\e[X_i^3f(X_i)^2]-\e[X_if(X_i)]\,\e[X_i^2f(X_i)]\\&=\mu_f(3,2,0)-\mu_f(1,1,0)\cdot\mu_f(2,1,0);\\
	\sigma_{35}&=CoV[X_if(X_i),\, X_i^2f'(X_i)]=\e[X_i^3f(X_i)f'(X_i)]-\e[X_if(X_i)]\,\e[X_i^2f'(X_i)]\\&=\mu_f(3,1,1)-\mu_f(1,1,0)\cdot\mu_f(2,0,1);\\
	\sigma_{44}&=CoV[X_i^2f(X_i),\, X_i^2f(X_i)]=\e[X_i^4f(X_i)^2]-\e[X_i^2f(X_i)]^2=\mu_f(4,2,0)-\mu_f^2(2,1,0);\\
	\sigma_{45}&=CoV[X_i^2f(X_i),\, X_i^2f'(X_i)]=\e[X_i^4f(X_i)f'(X_i)]-\e[X_i^2f(X_i)]\,\e[X_i^2f'(X_i)]\\&=\mu_f(4,1,1)-\mu_f(2,1,0)\cdot\mu_f(2,0,1);\\
	\sigma_{55}&=CoV[X_i^2f'(X_i),\, X_i^2f'(X_i)]=\e[X_i^4f'(X_i)^2]-\e[X_i^2f'(X_i)]^2=\mu_f(4,0,2)-\mu_f^2(2,0,1).
\end{align*}

%
\subsection{Proof of Theorem~\ref{Pele} (Sketch)}\label{Proof of Theorem Pele}
%
In what follows, we sketch the derivations for the variance and bias, respectively. 
Recall the abbreviation  $ \vartheta_f\coloneqq  \mu_f(2,1,0)-\mu^2\mu_f(0,1,0)$. 
First, we evaluate the gradient and Hessian of~$g$ in~$\bmu_Z$, which leads to $ \D $ and $ \Hes $ given by
\begin{align*}
	\D&=\frac{\lambda}{\vartheta_f}\Bigg(\frac{2 \mu_f(2,1,0)}{\mu},\, \mu ^2,\, 
	\frac{\mu ^2}{\lambda},\, -1,\frac{2\mu ^2}{\lambda} \Bigg)^\top, \\[2ex] 	
	\Hes&=\frac{\lambda}{\vartheta_f}\begin{pmatrix} 
		\frac{2 \mu_f(2,1,0) \left(\mu_f(2,1,0)+3\mu ^2 \mu_f(0,1,0)\right)}{\mu^2\vartheta_f} & * & * & * & * \\[1ex]
		\frac{4 \mu\cdot \mu_f(2,1,0)}{\vartheta_f} & \frac{2\mu^4}{\vartheta_f} & * & * & * \\[1ex]
		\frac{2\mu\cdot \mu_f(2,1,0)}{\lambda\vartheta_f} &  \frac{ \mu^4}{\lambda\vartheta_f} & 0 & * & * \\[1ex]
		-\frac{2\mu_f(2,1,0)+2\mu^2\mu_f(0,1,0)}{\mu\cdot\vartheta_f} & -\frac{ 2\mu^2}{\vartheta_f} & -\frac{\mu ^2}{\lambda\vartheta_f} & \frac{2}{\vartheta_f} & * \\[1ex]
		\frac{4\mu\cdot\mu_f(2,1,0)}{\lambda\vartheta_f} & \frac{2\mu^4}{\lambda\vartheta_f} & 0 & -\frac{2\mu ^2}{\lambda\vartheta_f} & 0
	\end{pmatrix}.
\end{align*}
Here, for the sake of readability, the upper triangle of the symmetric matrix~$\Hes$ was replaced by stars `$*$'. 
Applying the Delta method to Theorem~\ref{CLT_IG}, the asymptotic normality 
\begin{align*}
	\sqrt{n}\big(\hat{\lambda}_{f,\IG}-\lambda\big) \ \xrightarrow{\text{d}}\ \norm\big(0, \sigma^2\big)\  \text{with}\  \sigma^2=\sum_{i,j=1}^{5}d_id_j\sigma_{ij}
\end{align*}
follows. The 2nd-order Taylor approximation $ \hat{\lambda}_{f,\IG} - \lambda \approx \D(\overline{\Z}-\bmu_Z)+\tfrac{1}{2}(\overline{\Z}-\bmu_Z)^\top\Hes(\overline{\Z}-\bmu_Z)$ allows to conclude the asymptotic bias as
\begin{align*}
	\mathbb{B}_{f,\IG}&=\tfrac{1}{n} \Bigg( \tfrac{1}{2}\sum_{i,j=1}^{5}h_{ij}\sigma_{ij} \Bigg).
\end{align*}
The explicit expression for the asymptotic variance $\sigma_{f,\IG}^2 = \frac{\sigma^2}{n}$ follows by applying $ \D $ and Theorem \ref{CLT_IG}. After tedious calculations, we obtain the expression for~$\sigma_{f,\IG}^2$ stated in Theorem~\ref{Pele}.
Similarly, using~$\Hes$, the expression for the asymptotic bias is derived. 
This completes the proof of Theorem~\ref{Pele}.

%
\subsection{Proof of Corollary~\ref{korIG_MM} (Sketch)}\label{Proof_korIG_MM}
%
If $ f\equiv 1 $, we have $\mu_1(k,l,m)=\delta_{m,0}\,\e[X^k]$, recall \eqref{mu1klm}, where~$\delta_{\cdot,\cdot}$ denotes the Kronecker delta, 
so Theorem~\ref{Pele} simplifies considerably:
	\begin{align*}
		\sigma_{1,\IG}^2&=\frac{1}{n}\Bigg[\, \frac{\mu^4}{\vartheta_f^2}\bigg[\e[X^2]-\frac{\lambda \vartheta_f}{\mu}\bigg] 
		+\frac{2\lambda}{\vartheta_f^2}\bigg[\e[X^2]\Big(4\mu\e[X^2]-\lambda \vartheta_f\Big)-\mu^2\,\e[X^3]\bigg]
		\\[1ex]&\quad+\frac{\lambda^2}{\mu\vartheta_f^2}\bigg[ \mu^5+\mu^3\vartheta_f - \mu^3\cdot\e[X^2] -4\e[X^2]\,\e[X^3] +\mu\Big(3\e[X^2]^2+\e[X^4]\Big)\bigg]\, \Bigg],
	\end{align*}
where $ \vartheta_1 = \e[X^2]-\mu^2 = \V[X] = \mu^3/\lambda $, and
	\begin{align*}
		\mathbb{B}_{1,\IG}&=\frac{1}{n}\Bigg[\frac{1}{\mu\vartheta_f^2}\bigg[\mu^2\cdot\e[X^2]\big(\e[X^2]+3\mu^2\big)
		+\lambda\Big(\mu^5 +\mu\cdot\e[X^4]+2\e[X^2]\cdot\big(\mu^3-\e[X^3]\big)\Big)\bigg]
		\\[1ex]&\quad+\frac{\mu}{\vartheta_f^3}\bigg[
		\e[X^2]\Big( 2\e[X^2]^2	-\mu\cdot\e[X^3]+\mu^4\Big)-\mu^2\cdot\Big(\mu\, \e[X^3]+2\e[X^2]^2 +\mu^4\Big)
		\bigg]\Bigg].
	\end{align*}
Using the following moments \citep[see][p.~366]{Tweed57a},
\ba
\label{IGmom234}
\textstyle
\e[X^2]=\mu^2+\tfrac{\mu^3}{\lambda},
\quad \e[X^3]=\mu^3+\tfrac{3\mu^4}{\lambda}+\tfrac{3\mu^5}{\lambda^2},
\quad \e[X^4]=\mu^4+\tfrac{6\mu^5}{\lambda}+\tfrac{15\mu^6}{\lambda^2}+\tfrac{15\mu^7}{\lambda^3},
\ea
and after tedious calculations, this leads to
\begin{align*}
	\sigma_{1,\IG}^2&=\frac{1}{n}\Bigg[\mu\lambda+\frac{\lambda^3}{\mu}+\frac{\lambda^4}{\mu}+\frac{2\lambda^3}{\mu^5}\Bigg(4\mu^4\Big(1+\frac{\mu}{\lambda}\Big)^2 -\mu^4\Big(1+\frac{\mu}{\lambda}\Big)-\mu\Big(\mu^3+\frac{3\mu^4}{\lambda}+\frac{3\mu^5}{\lambda^2}\Big)	\Bigg)
	\\[1ex]&\quad+\frac{\lambda^4}{\mu^7}\Bigg(\frac{5\mu^6}{\lambda}+\frac{15\mu^7}{\lambda^2}+\frac{15\mu^8}{\lambda^3}+3\mu^5\Big(1+\frac{\mu}{\lambda}\Big)^2-4\mu^2\Big(1+\frac{\mu}{\lambda}\Big)\Big(\mu^3+\frac{3\mu^4}{\lambda}+\frac{3\mu^5}{\lambda^2}\Big)\Bigg)\Bigg]\\
	&=\cdots=\frac{2\lambda(\lambda+3\mu)}{n},
\end{align*}
and
\begin{align*}
	\mathbb{B}_{1,\IG}&=\frac{\lambda^2(4\mu+2\lambda)}{\mu^6}\Bigg[\frac{4\lambda^2\mu^5+14\lambda\mu^6+15\mu^7}{\lambda^2(4\lambda+\mu)}-\frac{\mu^5}{\lambda} -\frac{3\mu^6}{\lambda^2}\Bigg] =\cdots=\frac{3(\lambda+3\mu)}{n}.
\end{align*}
So the proof of Corollary~\ref{korIG_MM} is complete.

%
\subsection{Proof of Corollary~\ref{korIG_ML} (Sketch)}\label{Proof_korIG_ML}
%
Recall from \eqref{mu1Xklm} that for $ f(x)=1/x $, we have $\mu_{1/x}(k,l,m)=(-1)^m\,\e[X^{k-l-2m}]$. 
In particular, $ \vartheta_f = \e[X]-\mu^2\,\e[X^{-1}] = \mu-\mu^2\,(\mu^{-1}+\lambda^{-1}) =-\mu^2/\lambda $.
Furthermore, positive and negative moments are related to each other by \eqref{MomIG}, where closed-form expressions for moments of order 2--4 are given by \eqref{IGmom234}. This can be used to simplify the joint moments involved in Theorem~\ref{Pele} as follows:
\begin{align*}
\mu_{1/x}(0,1,0) = {} & \e[X^{-1}] = \tfrac{1}{\mu}+\tfrac{1}{\lambda}, && 
\mu_{1/x}(3,0,1) = -\e[X]=-\mu, \\
\mu_{1/x}(0,2,0) = {} & \e[X^{-2}]
 = \tfrac{1}{\mu^{5}}\,(\mu^3+\tfrac{3\mu^4}{\lambda}+\tfrac{3\mu^5}{\lambda^2}), && 
\mu_{1/x}(3,1,0) = \e[X^{2}]=\mu^2+\tfrac{\mu^3}{\lambda}, \\
\mu_{1/x}(1,1,0) = {} & 1, && 
\mu_{1/x}(3,1,1) = -1, \\
\mu_{1/x}(1,2,0) = {} & \e[X^{-1}] = \tfrac{1}{\mu}+\tfrac{1}{\lambda}, && 
\mu_{1/x}(3,2,0) = \e[X]=\mu, \\
\mu_{1/x}(2,0,1) = {} & -1, && 
\mu_{1/x}(4,0,2) = 1, \\
\mu_{1/x}(2,1,0) = {} & \e[X]=\mu, && 
\mu_{1/x}(4,1,1) = -\e[X]=-\mu, \\
\mu_{1/x}(2,1,1) = {} & -\e[X^{-1}] = -\tfrac{1}{\mu}-\tfrac{1}{\lambda}, && 
\mu_{1/x}(4,2,0) = \e[X^{2}]=\mu^2+\tfrac{\mu^3}{\lambda}, \\
\mu_{1/x}(2,2,0) = {} & 1. 
\end{align*}
After tedious calculations, Theorem~\ref{Pele} simplifies to
\begin{align*}
	\sigma_{1/x,\IG}^2&=\frac{1}{n}\Bigg[\frac{2\lambda^3}{\mu^3}\Big(\mu\big(6\mu-4\mu+\mu\big)-\mu\big(\mu-2\mu\big) \Big)+\frac{\lambda^4}{\mu^5}\bigg(\mu^3+\tfrac{3\mu^4}{\lambda}+\tfrac{3\mu^5}{\lambda^2}
	\\[1ex]&\quad+\mu^2(1+\tfrac{\mu}{\lambda})\big(\mu-\tfrac{\mu^2}{\lambda}\big) +2\mu^3+\mu\Big(3\mu^2+\mu^2(1+\tfrac{\mu}{\lambda})\Big)\bigg)\Bigg]
	=\cdots=\frac{2\lambda^2}{n},
\end{align*}
for the variance, and to
\begin{align*}
	\mathbb{B}_{1/x,\IG}&=\frac{\lambda^2}{\mu^5}\Bigg[4\mu^4+\tfrac{3\mu^5}{\lambda}+\lambda\Big(3\mu^3-\mu^3\big(1+\tfrac{\mu}{\lambda}\big)+\tfrac{3\mu^4}{\lambda}+\tfrac{3\mu^5}{\lambda^2}\Big)\Bigg]
	\\[1ex]&\quad-\frac{\lambda^3}{\mu^5}\Bigg[\mu\big(1+\tfrac{\mu}{\lambda}\big)\Big(\mu^2+3\mu^2\big(1+\tfrac{\mu}{\lambda}\big)\Big)-\mu^3-\mu^3\big(1+\tfrac{\mu}{\lambda}\big)\Bigg]
	=\cdots=\frac{3\lambda}{n}.
\end{align*}
for the bias. 
So the proof of Corollary~\ref{korIG_ML} is complete.

%
\subsection{Proof of Theorem~\ref{CLT_Nb}}\label{Proof of Theorem CLT_Nb}
%
The asymptotic normality immediately follows from the Lindeberg--L\'evy CLT \citep[p.~28]{serfling80}. 
For the covariances, we get 
\begin{align*}
	\sigma_{11}&=CoV[X_i,\, X_i]=\sigma^2;\\
	\sigma_{12}&=CoV[X_i,\, f(X_i+1)]=\e[X_if(X_i+1)]-\e[X_i]\,\e[f(X_i+1)]\\
	&=\tilde{\mu}_f(1,0,1)-\mu\cdot\tilde{\mu}_f(0,0,1);\\
	\sigma_{13}&=CoV[X_i,\, X_if(X_i)]=\e[X_i^2f(X_i)]-\e[X_i]\,\e[X_if(X_i)]\\
	&=\tilde{\mu}_f(2,1,0)-\mu\cdot\tilde{\mu}_f(1,1,0);\\
	\sigma_{14}&=CoV[X_i,\, X_if(X_i+1)]=\e[X_i^2f(X_i+1)]-\e[X_i]\,\e[X_if(X_i+1)]\\
	&=\tilde{\mu}_f(2,0,1)-\mu\cdot\tilde{\mu}_f(1,0,1);\\
	\sigma_{22}&=CoV[f(X_i+1),\, f(X_i+1)]=\e[f(X_i+1)^2]-\e[f(X_i+1)]^2\\
	&=\tilde{\mu}_f(0,0,2)-\tilde{\mu}_f^2(0,0,1);\\
	\sigma_{23}&=CoV[f(X_i+1),\, X_if(X_i)]=\e[X_if(X_i)f(X_i+1)]-\e[f(X_i+1)]\,\e[X_if(X_i)]\\
	&=\tilde{\mu}_f(1,1,1)-\tilde{\mu}_f(0,0,1)\cdot\tilde{\mu}_f(1,1,0);\\
	\sigma_{24}&=CoV[f(X_i+1),\, X_if(X_i+1)]=\e[X_if(X_i+1)^2]-\e[f(X_i)]\,\e[X_if(X_i+1)]\\
	&=\tilde{\mu}_f(1,0,2)-\tilde{\mu}_f(0,0,1)\cdot\tilde{\mu}_f(1,0,1);\\
	\sigma_{33}&=CoV[X_if(X_i),\, X_if(X_i)]=\e[X_i^2f(X_i)^2]-\e[X_if(X_i)]^2=\tilde{\mu}_f(2,2,0)-\tilde{\mu}_f^2(1,1,0);\\
	\sigma_{34}&=CoV[X_if(X_i),\, X_if(X_i+1)]=\e[X_i^2f(X_i)f(X_i+1)]-\e[X_if(X_i)]\,\e[X_if(X_i+1)]\\&=\tilde{\mu}_f(2,1,1)-\tilde{\mu}_f(1,1,0)\cdot\tilde{\mu}_f(1,0,1);\\
	\sigma_{44}&=CoV[X_if(X_i+1),\, X_if(X_i+1)]=\e[X_i^2f(X_i)^2]-\e[X_if(X_i)]^2\\
	&=\tilde{\mu}_f(2,0,2)-\tilde{\mu}_f^2(1,0,1).
\end{align*}

\subsection{Proof of Theorem~\ref{Ansu} (Sketch)}\label{Proof of Theorem Ansu}
%
In what follows, we sketch the derivations for the variance and bias, respectively. 
Recall the abbreviation  $ \eta_1\coloneqq  \tilde{\mu}_f(1,1,0)-\mu\cdot\tilde{\mu}_f(0,0,1)$. 
Furthermore, in order to denote the results more compactly, let us abbreviate $\tilde{\mu}_{klm} := \tilde{\mu}_f(k,l,m)$. 
First, we evaluate the gradient and Hessian of~$g_\nu$ in~$\bmu_Z$, which leads to $ \D $ and $ \Hes $ given by
\begin{align*}
	\D&=\tfrac{1}{\eta_1^2} \Big(\tilde{\mu}_{110}\, (\tilde{\mu}_{101} - \tilde{\mu}_{110}),\ \mu^2\, (\tilde{\mu}_{101} - \tilde{\mu}_{110}),\ \mu \, (\mu\, \tilde{\mu}_{001} - \tilde{\mu}_{101}),\ \mu\, (\tilde{\mu}_{110}-\mu\, \tilde{\mu}_{001}) \Big),
\end{align*}
\scriptsize
\begin{align*}
		\Hes&=\frac{1}{\eta_1^3}\begin{pmatrix} 
			2 \tilde{\mu}_{110} \tilde{\mu}_{001} (\tilde{\mu}_{101} - \tilde{\mu}_{110}) & * & * & * \\[1ex]
 2 \mu\, \tilde{\mu}_{110}\, (\tilde{\mu}_{101} - \tilde{\mu}_{110}) & 2 \mu^3\, (\tilde{\mu}_{101} - \tilde{\mu}_{110}) & * & * \\[1ex]
\mu\, \tilde{\mu}_{001}\, (2 \tilde{\mu}_{110} - \tilde{\mu}_{101}) - \tilde{\mu}_{101}\, \tilde{\mu}_{110} & \mu^2\, (\mu\, \tilde{\mu}_{001} - 2 \tilde{\mu}_{101} + \tilde{\mu}_{110}) & 2 \mu\, (\tilde{\mu}_{101}-\mu\, \tilde{\mu}_{001}) & * \\[1ex]
 \tilde{\mu}_{110}\, (\tilde{\mu}_{110}-\mu\, \tilde{\mu}_{001}) & \mu^2\, (\tilde{\mu}_{110}-\mu\, \tilde{\mu}_{001}) & \mu\, (\mu\, \tilde{\mu}_{001} - \tilde{\mu}_{110}) & 0
		\end{pmatrix}.
	\end{align*}
	\normalsize
	Here, for the sake of readability, the upper triangle of the symmetric matrix~$\Hes$ was replaced by stars `$*$'. 
	Applying the Delta method to Theorem~\ref{CLT_Nb}, the asymptotic normality 
$$
\textstyle
		\sqrt{n}\big(\hat{\nu}_{f,\nb}-\nu\big) \ \xrightarrow{\text{d}}\ \norm\big(0, \sigma^2\big)\  \text{with}\  \sigma^2=\sum_{i,j=1}^{4}d_id_j\sigma_{ij}
$$
	follows. The 2nd-order Taylor approximation $ \hat{\nu}_{f,\nb} - \nu \approx \D(\overline{\Z}-\bmu_Z)+\tfrac{1}{2}(\overline{\Z}-\bmu_Z)^\top\Hes(\overline{\Z}-\bmu_Z)$ allows to conclude the asymptotic bias as
$$
\textstyle
		\mathbb{B}_{f,\nb}\ =\ \tfrac{1}{n} \Bigg( \tfrac{1}{2}\sum_{i,j=1}^{4}h_{ij}\sigma_{ij} \Bigg).
$$
	The explicit expressions for the asymptotic variance $\sigma_{f,\nb}^2 = \frac{\sigma^2}{n}$ follow by applying $ \D $ and Theorem \ref{CLT_Nb}. After tedious calculations, we obtain the expressions stated in Theorem~\ref{Ansu}, which completes the proof.

%
\subsection{Proof of Theorem~\ref{Fati} (Sketch)}\label{Proof of Theorem Fati}
%
In what follows, we sketch the derivations for the variance and bias, respectively. 
Recall the abbreviation  $ \eta_2\coloneqq  \tilde{\mu}_f(1,0,1)-\mu\cdot\tilde{\mu}_f(0,0,1)$, and let us again the abbreviations $\tilde{\mu}_{klm} := \tilde{\mu}_f(k,l,m)$.
First, we evaluate the gradient and Hessian of~$g_\pi$ in~$\bmu_Z$, which leads to $ \D $ and $ \Hes $ given by
\begin{align*}
	\D&=\tfrac{1}{\eta_2^2} \Big(\tilde{\mu}_{001}\, (\tilde{\mu}_{101} - \tilde{\mu}_{110}),\ \mu\, (\tilde{\mu}_{101} - \tilde{\mu}_{110}),\ \mu \, \tilde{\mu}_{001} - \tilde{\mu}_{101},\ \tilde{\mu}_{110}-\mu\, \tilde{\mu}_{001} \Big),
\end{align*}
\scriptsize
\begin{align*}
	\Hes&=\frac{1}{\eta_2^3}\begin{pmatrix} 
		  2 \tilde{\mu}_{001}^2\, (\tilde{\mu}_{101} - \tilde{\mu}_{110}) & * & * & * \\[1ex]
(\mu\, \tilde{\mu}_{001} + \tilde{\mu}_{101}) (\tilde{\mu}_{101} - \tilde{\mu}_{110}) & 2 \mu^2\, (\tilde{\mu}_{101} - \tilde{\mu}_{110}) & * & * \\[1ex]
\tilde{\mu}_{001}\, (\mu\, \tilde{\mu}_{001} - \tilde{\mu}_{101}) & \mu\, (\mu\, \tilde{\mu}_{001} - \tilde{\mu}_{101}) & 0 & * \\[1ex]
\tilde{\mu}_{001}\, (2 \tilde{\mu}_{110} - \mu\, \tilde{\mu}_{001} - \tilde{\mu}_{101}) & \mu\, (2 \tilde{\mu}_{110} - \mu\, \tilde{\mu}_{001} - \tilde{\mu}_{101}) & \tilde{\mu}_{101} -\mu\, \tilde{\mu}_{001} & 2 \mu\, \tilde{\mu}_{001} - 2 \tilde{\mu}_{110}
	\end{pmatrix}.
\end{align*}
\normalsize 
Here, for the sake of readability, the upper triangle of the symmetric matrix~$\Hes$ was replaced by stars `$*$'. 
Then, the remaining steps are like in Appendix~\ref{Proof of Theorem Ansu}. This completes the proof of Theorem~\ref{Fati}.

\subsection{Proof of Lemma~\ref{facZx}}
\label{Proof of Lemma facZx}
We proof Lemma~\ref{facZx} by induction.
	
	\smallskip
	\textbf{Base case:}\quad
	For $k=0$ with $x_{(0)}=1$, we get $\e[z^X]=\pgf(z)$.
	
	\smallskip
	\textbf{Inductive step:}\quad
	Let us assume that the statement holds for a given $k$. Then,
	$$
		\e[X_{(k+1)}z^X]\ =\ z^{k+1}\, \e[X_{(k+1)}z^{X-k-1}]=z^{k+1}\, \frac{d}{dz}\, \e[X_{(k)}z^{X-k}].
	$$
	By the induction hypothesis, we get
		\begin{align*}
		\e[X_{(k+1)}z^X]
		&=z^{k+1}\,  \frac{d}{dz}\, \frac{(1-\pi)^k (\nu+k-1)_{(k)}}{(1-(1-\pi)z)^k}\, \bigg(\frac{\pi}{1-(1-\pi)z} \bigg)^\nu\\
		&=z^{k+1}\, (1-\pi)^k (\nu+k-1)_{(k)}\pi^\nu\cdot  \frac{d}{dz}\,  (1-(1-\pi)z)^{-(\nu+k)} \\
		&=z^{k+1}\, (1-\pi)^k (\nu+k-1)_{(k)}\pi^\nu\cdot  (\nu+k)(1-\pi)\,  (1-(1-\pi)z)^{-(\nu+k+1)} \\
		&=z^{k+1}\frac{(1-\pi)^{k+1}(\nu+k)_{(k+1)}}{(1-(1-\pi)z)^{k+1}}\, \pgf(z).
	\end{align*}
	This completes the proof of Lemma~\ref{facZx}.

\end{document}